\documentclass[11pt, A4]{article}\pdfoutput=1

\usepackage{graphicx}
\usepackage{epstopdf}
\usepackage{amsmath}
\usepackage{amssymb}
\usepackage{amsfonts}
\usepackage{amsthm}
\usepackage{mathrsfs}
\usepackage{ccaption}
\usepackage[usenames]{color}
\usepackage{comment}
\usepackage{overpic}
\usepackage{array}

\definecolor{AV}{rgb}{0.65,0.0,0}

\usepackage[
      colorlinks=true,
      linkcolor=blue,
      urlcolor=blue,
      filecolor=blue,
      citecolor=red,
      pdfstartview=FitV,
      pdftitle={},
      pdfauthor={},
      pdfsubject={},
      pdfkeywords={},
      pdfpagemode=None,
      bookmarksopen=true
]{hyperref}

\usepackage{graphicx}
\usepackage{epstopdf}
\usepackage{epsfig,amssymb}

\newcommand{\rom}[1]{\mathrm{#1}}

\newcommand{\beq}{\begin{equation}}
\newcommand{\eeq}{\end{equation}}
\newcommand{\be}{\begin{equation}}
\newcommand{\ee}{\end{equation}}
\newcommand{\beqa}{\begin{eqnarray}}
\newcommand{\eeqa}{\end{eqnarray}}
\newcommand{\beqar}{\begin{eqnarray*}}
\newcommand{\eeqar}{\end{eqnarray*}}
\newcommand{\bea}{\begin{eqnarray}}
\newcommand{\eea}{\end{eqnarray}}

\newcommand{\nn}{\nonumber}
\def\cV{\mathcal{V}}
\def \RR{{\mathbb{R}}}

\def\g{\mathfrak{g}}

\def\su{\mathfrak{su}}

\def\sl{\mathfrak{sl}}
\def\gl{\mathfrak{gl}}

\def\cA{\mathcal{A}}

\def\cK{\mathcal{K}}
\def\cL{\mathcal{L}}
\def\cM{\mathcal{M}}
\def\cN{\mathcal{N}}

\def\cQ{\mathcal{Q}}

\def\cV{\mathcal{V}}

\def\beq{\begin{eqnarray}}
\def\eeq{\end{eqnarray}}
\def\mf{\mathfrak}
\def\Aone{B_2}
\def\Atwo{\omega_3}
\def\Azero{\chi_1}

\def\axone{\chi_2}
\def\axtwo{\chi_3}

\def\gauge{B_1}

\def\nn{\nonumber}
\def\half{\frac{1}{2}}

\begin{document}

\begin{flushright}
 {\tt ULB-TH/10-10}
\end{flushright}
\vskip 1cm

\begin{center}
{\bf \Large Extremal solutions of the S$^3$ model\\ \vspace{.1cm} and
nilpotent orbits of $\mathrm{G}_{2(2)}$}
\end{center}

\vspace{0.75cm}

\begin{center}
{\bf \large Sung-Soo Kim, Josef Lindman H\"ornlund,\\Jakob Palmkvist and Amitabh Virmani}

\vspace{4mm} { \normalsize
{\em Physique Th\'eorique et Math\'ematique \\  Universit\'e Libre de
    Bruxelles\\and\\ International Solvay Institutes\\ Campus
    Plaine C.P. 231, B-1050 Bruxelles,  Belgium \\
}

\vspace*{0.5cm}

\texttt{sungsoo.kim, jlindman, jakob.palmkvist,  avirmani@ulb.ac.be}}

\end{center}\vspace{0.5cm}


\begin{abstract}
We study extremal black hole solutions of the S$^3$ model (obtained by setting S=T=U in the STU model) using group theoretical methods. Upon dimensional reduction over time, the S$^3$ model exhibits the pseudo-Riemannian coset structure $G/{\tilde K}$ with $G= \mathrm{G}_{2(2)}$ and $\tilde K = \mathrm{SO}_0(2,2)$.  We study nilpotent $\tilde K$-orbits of $\mathrm{G}_{2(2)}$ corresponding to non-rotating single-center extremal solutions. We find six such distinct $\tilde K$-orbits.  Three of these orbits are supersymmetric, one is non-supersymmetric, and two are unphysical.  We write general solutions and discuss examples in all four physical orbits. We show that all solutions in supersymmetric orbits when uplifted to five-dimensional minimal supergravity have single-center Gibbons-Hawking space as their four-dimensional Euclidean hyper-K\"ahler base space. We construct hitherto unknown extremal (supersymmetric as well as non-supersymmetric) pressureless black strings of minimal five-dimensional supergravity and briefly discuss their relation to black rings.
\end{abstract}

\newpage

\vspace{0.5cm} \tableofcontents

\newpage

\setlength{\unitlength}{1mm}


\section{Introduction and summary}
\label{sec:Introduction}

Supersymmetric black hole solutions of various {four and five-dimensional supergravity theories are fairly} well understood, so much so that in some cases a complete classification is available. More recently the focus of investigation has shifted to extremal but not necessarily supersymmetric black holes. There exists a large body of literature exploring extremal black holes in various supergravity theories starting with the observation that not only BPS black holes \cite{Ferrara:1995ih, Ferrara:1996um} but also certain non-BPS extremal black holes show attractor behavior \cite{Ferrara:1997tw, Sen:2005wa, Goldstein:2005hq}.

The attractor behavior of extremal but not supersymmetric solutions was further developed in \cite{Ceresole:2007wx, Andrianopoli:2007gt,Cardoso:2007ky, Perz:2008kh}, where it was observed that  the second order flow equations underlying extremal solutions can be factorized  into first order equations using a `fake superpotential'. Using the approach of the fake superpotential a large class of extremal solutions has been constructed and analyzed in \cite{Hotta:2007wz, Gimon:2007mh, Bellucci:2008sv, Gimon:2009gk}. A systematic construction of fake superpotentials for non-BPS extremal solutions was given in \cite{Bossard:2009we} using group theoretical techniques developed in \cite{Bossard:2009at, Bossard:2009my}. The  considerations of  \cite{Bossard:2009we, Bossard:2009at, Bossard:2009my} are sufficiently general and powerful for providing a complete classification of extremal solutions in all supergravity theories in which scalars parameterize a symmetric moduli space.

In this paper we use the techniques of \cite{Bossard:2009we, Bossard:2009at, Bossard:2009my} to study extremal solutions of minimal $N=2, D=4$  supergravity coupled to one vector multiplet with cubic prepotential. This
theory is rich enough to harbor many extremal solutions of interest and simple enough to lend to a completely explicit and systematic analysis in the approach of \cite{Bossard:2009we, Bossard:2009at, Bossard:2009my}. Perhaps the simplest way to obtain this theory is via circle reduction of minimal five-dimensional supergravity. The theory is  known as the S$^3$ model as its prepotential is cubic in the modulus of the theory. It can also be obtained by setting the three moduli equal in the STU model.

The approach of \cite{Bossard:2009we, Bossard:2009at, Bossard:2009my} does not rely on supersymmetry and treats all extremal solutions on the same footing. It is based on the pseudo-Riemannian non-linear sigma models obtained by reducing a theory to three spatial dimensions. It captures all static spherically symmetric solutions of all gravity theories that upon dimensional reduction give rise to pseudo-Riemannian non-linear sigma models based on a coset $G/\tilde{K}$, where the group $\tilde{K}$ is  in general \emph{non-compact}. In particular the approach of \cite{Bossard:2009we, Bossard:2009at, Bossard:2009my} is applicable to our theory as upon dimensional reduction over time the S$^3$ model gives rise to a $\mathrm{G}_{2(2)}/ \mathrm{SO}_0(2,2)$ sigma model. This sigma model is pseudo-Riemannian because $\mathrm{SO}_0(2,2)$ is not the maximal compact subgroup of $\mathrm{G}_{2(2)}$.

The key point of \cite{Bossard:2009we, Bossard:2009at, Bossard:2009my} is to identify the
charge matrix of an extremal solution as an element of a nilpotent orbit of the three-dimensional duality group. The observation of nilpotency of charge matrices goes back to  \cite{Gunaydin:2005mx}, where it was first observed that the charge matrix of four-dimensional supersymmetric black holes must be nilpotent. These ideas were further developed in \cite{Berkooz:2008rj, Bossard:2009at, Bossard:2009my}. In particular, in \cite{Bossard:2009at} it was argued that nilpotency of the charge matrix holds in general for an extremal solution that can be obtained as a limit of a non-extremal solution. It was concluded that by studying certain nilpotent $\tilde{K}$-orbits of the Lie algebra of $G$, one can completely classify extremal solutions of a theory.

The aim of this paper is to carry out a detailed and explicit analysis of nilpotent orbits for the S$^3$ model. A study of nilpotent charge matrices for this theory was also done in \cite{Gaiotto:2007ag, Berkooz:2008rj, Bergshoeff:2008be}. In \cite{Gaiotto:2007ag} it was noted that nilpotent charge matrices are directly related to the attractor behavior. Two distinct attractor flows were constructed, one BPS and one non-BPS. In \cite{Berkooz:2008rj}
nilpotency of the charge matrices was checked for certain solutions of five-dimensional minimal supergravity. We revisit the analysis of \cite{Gaiotto:2007ag, Berkooz:2008rj}  using recently developed group theoretic techniques \cite{Bossard:2009we, Bossard:2009at, Bossard:2009my}. Our main results are  summarized as follows:
\begin{itemize}
\item  We study nilpotent $\tilde{K}$-orbits of $\tilde{\mathfrak{p}}$ for the S$^3$ model, where $\tilde{\mathfrak{p}}$ is the complement of the Lie algebra of $\tilde{K}$ in $\mathfrak{g}_{2(2)}$ with respect to the Killing form.  We find six orbits of extremal black holes that can be obtained as a limit of non-extremal black holes.  We write explicit expressions for scalar and electromagnetic charges for each orbit.

\item We show that two of six orbits are unphysical. Among the rest, three are supersymmetric, and one is non-supersymmetric. Out of the three supersymmetric orbits only one orbit corresponds to extremal black holes with  non-zero horizon area. This supersymmetric orbit and the non-supersymmetric orbit precisely correspond to the BPS and non-BPS attractor flows of \cite{Gaiotto:2007ag}. We discuss new and known examples of extremal black holes in all four physical orbits.

\item We connect the classification of supersymmetric black holes in terms of the $\tilde{K}$-orbits with the analysis of \cite{Gauntlett:2002nw}. In particular, we show that solutions corresponding to all three supersymmetric orbits, when uplifted to five-dimensional minimal supergravity, have single-center Gibbons-Hawking space as their four-dimensional Euclidean hyper-K\"ahler base space. We also show the converse, namely, all static
extremal asymptotically flat black holes that can be obtained via dimensional reduction of the single-center supersymmetric Gibbons-Hawking form are
supersymmetric and belong to one of the supersymmetric orbits. Our analysis hence provides a partial proof of the conjecture \cite{Gauntlett:2002nw} that dimensional reduction of solutions of five-dimensional minimal supergravity with Gibbons-Hawking base gives the entire timelike class of supersymmetric solutions of the S$^3$ model.

\item We construct a three-parameter family of supersymmetric  black strings with independent $\rom{M2^3-M5^3-P}$ charges. Its macroscopic entropy can be reproduced \cite{Larsen:2005qr}
from the Maldacena-Strominger-Witten CFT \cite{Maldacena:1997de}. This family contains two distinct two parameter sub-families of pressureless black strings. One of these sub-families  is well known \cite{Bena:2004wv} and  describes the infinite radius limit of the supersymmetric black ring \cite{Elvang:2004rt}. The second two parameter sub-family has not been previously
discussed in the literature. We discuss its relation to black rings.
\item We also construct a three parameter family of \emph{non-supersymmetric} black strings with independent $\rom{M2^3-M5^3-P}$ charges. This family contains a  two parameter sub-family of pressureless black strings.
 \end{itemize}

The motivation behind studying pressureless black strings comes from the fact that all known smooth black rings with charges \cite{Elvang:2004rt, Elvang:2003yy} and dipoles \cite{Emparan:2004wy} become pressureless strings in the infinite radius limit. The connection between pressureless black strings and black rings was first made explicit in \cite{Elvang:2003mj}. In the \emph{blackfold} approach \cite{Emparan:2009cs} it appears that all pressureless black strings describe the infinite radius limit of some black ring.

The rest of the paper is organized as follows.  In  section \ref{sec:NilpotentOrbits}, we start with an overview of the approach of nilpotent orbits that underlies our study. To set notation
we give a  brief review of the S$^3$ model in section \ref{sec:S3theory}. The Lagrangian description of the S$^3$ model is presented as the circle reduction of five-dimensional minimal supergravity and is related to its $N=2$ prepotential description. The dimensional reduction over time from four to three dimensions is performed in section \ref{sec:reductionontime}. The five-dimensional uplift to minimal supergravity and some basic facts about supersymmetric solutions of five-dimensional minimal supergravity are collected in section \ref{sec:timereduction}. In the next section nilpotent orbits of $\mathrm{G}_{2(2)}$ are studied. After reviewing generalities of $\mathrm{G}_{2(2)}$ (section \ref{genG2}), nilpotent orbits of the complex $\mathrm{G}_2$, nilpotent orbits of the split real form $\mathrm{G}_{2(2)}$, and finally nilpotent $\tilde K$-orbits of  $\mathrm{G}_{2(2)}$ are analyzed in sections \ref{sec:orbitstructure}, \ref{sec:orbitstructure1}, \ref{sec:orbitstructure2}, respectively. Supersymmetric orbits are discussed in more detail in section \ref{SupersymmetricOrbits}
 and the non-supersymmetric orbit is discussed in section \ref{NonSupersymmetricOrbits}. For each orbit we present general expressions for charges and discuss examples. The three-parameter families of supersymmetric and non-supersymmetric black strings with independent $\rom{M2^3-M5^3-P}$ charges are given in sections \ref{sec:susystring} and \ref{sec:nonsusystring}, respectively.
In section \ref{sec:extremalsolutions} we discuss how our analysis fits in with the approach of
\cite{Gauntlett:2002nw}.  Details of the coset model construction are relegated to the appendix.


\setcounter{equation}{0}
\section{Nilpotent orbits}
\label{sec:NilpotentOrbits}

Since the work of Breitenlohner, Maison, and Gibbons \cite{Breitenlohner:1987dg} it is known that spherically symmetric black holes for a wide class of four-dimensional gravity theories correspond to geodesic segments on coset manifolds $G/{\tilde K}$. The coset manifold is the target space of a three-dimensional sigma model constructed from the four-dimensional gravity theory by performing a dimensional reduction over time and dualizing the resulting vectors into scalars. The group $G$ is the duality group of the scalars in three dimensions and $\tilde K$ a certain non-compact subgroup of $G$.  A geodesic on the coset manifold is completely specified by its starting point $p \in G/{\tilde K}$ and its velocity at $p$. The velocity at the point $p$ is the conserved Noether charge $\mathcal{Q} \in \mathfrak{g}$ taking values in the Lie algebra $\mathfrak{g}$ of $G$. From the four-dimensional spacetime point of view the starting position $p$ of the geodesic is associated with the values of the moduli at spatial infinity and the velocity $\mathcal{Q}$ at the point $p$ is associated with the four-dimensional conserved charges.

The action of $G$ on a given solution is such that it acts both on the position and the velocity of the corresponding geodesic.  The subgroup of $G$ that keeps the starting point $p$ fixed is $\tilde K$. The subgroup $\tilde K$ thus generates the full set of transformations of the conserved charge $\cQ$. It was shown in \cite{Breitenlohner:1987dg} that using $\tilde K$ one can generate the full class of single-center non-extremal spherically symmetric black holes of a given theory starting from the Schwarzschild black hole. In other words, all single-center non-extremal spherically symmetric black holes lie in a single
$\tilde K$-orbit containing the Schwarzschild black hole. It was also shown in \cite{Breitenlohner:1987dg} that the extremality parameter $c$ of a black hole corresponds to the `speed' of the geodesic,  i.e., $c^2 = \frac{1}{4}\mathrm{tr}(\mathcal{Q}^2).$ Since the reduction to three dimensions is performed over time, the resulting coset $G/{\tilde K}$ is pseudo-Riemannian. Thus, there are also null geodesics on the coset manifold.  When the extremality parameter goes to zero, the geodesic becomes null and the corresponding black hole becomes extremal.

The fact that $\tilde K$ preserves the point $p$ induces an action of $\tilde K$ on the tangent space $T_p(G/\tilde{K})$, and this in turn induces the reductive decomposition
\begin{equation}
\mathfrak{g} = \tilde{\mathfrak{k}} \oplus \tilde{\mathfrak{p}},
\end{equation}
where $\tilde{\mathfrak{k}}$ is the Lie algebra of $\tilde K$ and $\tilde{\mathfrak{p}}$ is isomorphic to $T_p(G/{\tilde K})$ via the standard isomorphism $T_pG \cong \mathfrak{g}$. These vector spaces hence obey the relation
\begin{equation}
[\tilde{\mathfrak{k}}, \tilde{\mathfrak{p}}] \subset \tilde{\mathfrak{p}}.
\end{equation}
If we assume  asymptotic flatness and that the geodesic corresponding to a given black hole starts at the identity coset, then the geodesics is given by
\begin{equation}\label{M}
\cM = \exp \left(-\frac{1}{r} \mathcal{Q}\right),
\end{equation}
where $\mathcal{Q} \in \tilde{\mathfrak{p}}$ {and $r$ is a radial coordinate.} However, not all geodesics on the coset manifold lead to regular bona-fide black holes. Already in their investigation of vacuum five-dimensional gravity, for which $G =\mathrm{SL}(3,\mathbb{R})$ and $\tilde{K} = \mathrm{SO}(2,1)$, Maison and Dobiasch \cite{Dobiasch:1981vh} observed that one could impose the condition that the charge matrix $\mathcal{Q} \in \mathfrak{sl}(3, \mathbb{R})$  should have vanishing determinant,
\begin{equation}
\det \mathcal{Q} = 0,
\end{equation}
for the black hole to be regular outside the horizon. If we further restrict ourselves to extremal black holes, i.e., $c=0$,  the conserved charge $\cQ$ becomes nilpotent,
\begin{equation}
\label{eqn:thirddegreenilpotency}
\mathcal{Q}^3 = 0,
\end{equation}
because the two invariant polynomials of $\mathfrak{sl}(3, \mathbb{R})$, $\det \mathcal{Q}$ and $\mathrm{tr}(\mathcal{Q}^2)$, now vanish.

In fact, nilpotency of the charge matrix seems to be a generic requirement for extremal black holes of all theories discussed in \cite{Breitenlohner:1987dg}. For supersymmetric black holes of $N \ge 2$ supergravity theories it was shown in \cite{Gunaydin:2005mx} that supersymmetry requires the charge matrix $\cQ$ to be nilpotent. In \cite{Bossard:2009at} it was  observed that the charge matrix $\mathcal{Q}$ for any asymptotically flat non-extremal axisymmetric solution satisfies\footnote{This is true for all cases in the classification of \cite{Breitenlohner:1987dg} except for the two cases involving real forms of $\mathrm{E}_8$.}
\begin{equation}
\mathcal{Q}^3 - \frac{1}{4}\mathrm{tr}(\mathcal{Q}^2) \mathcal{Q} = 0.
\end{equation}
Therefore, it follows that a non-rotating extremal solution that can be obtained as an extremal limit of a non-extremal solution is characterized by a nilpotent charge matrix $\cQ$.
Equivalently, by classifying nilpotent $\tilde K$-orbits of $\tilde {\mf{ p}}$,
one obtains a complete classification of single center extremal black holes \cite{Bossard:2009we, Bossard:2009at, Bossard:2009my}. Note anyhow that nilpotency is not a sufficient condition for a black hole to be regular everywhere outside  the horizon.


\section{Single modulus S$^3$ $N=2, D=4$ supergravity}
In this section we start by describing the $N=2, D=4$ single modulus S$^3$ model. We define the electromagnetic charges in this model and discuss the uplift to minimal five-dimensional supergravity.
\subsection{The theory}
\label{sec:S3theory}

The single modulus S$^3$ model in four dimensions consists of the $N=2$ gravity multiplet coupled to a vector multiplet. It is the consistent truncation of the STU model \cite{Duff:1995sm} where the $(S, T, U)$ moduli are identified with each other. The S$^3$ model can also be obtained from the circle reduction of five-dimensional minimal supergravity to four dimensions.
In this section we first present the Lagrangian description of the model obtained directly from the circle reduction of five-dimensional minimal supergravity. This is the description that we mostly use in this paper. For completeness we also present the $N=2$ prepotential construction and show that it is equivalent to the circle reduction of five-dimensional minimal supergravity.

The bosonic sector of five-dimensional minimal supergravity contains a metric $g_5$ and a Maxwell potential $A$ whose field strength is $F= dA$. The bosonic part of the Lagrangian takes the form of the Einstein-Maxwell theory with a Chern-Simons term,
\begin{equation}
\mathcal{L}_5 = R_5 \star_5  \mathbf{1} - \half\star_5  F \wedge F +\frac{1}{3\sqrt 3} F \wedge F \wedge A. \label{5dsugra}
\end{equation}
To perform the dimensional reduction to four dimensions, we assume that the extra spatial direction (denoted by $z$) is compact and a Killing direction in the five-dimensional spacetime. Using the standard Kaluza-Klein ansatz to yield a four-dimensional Lagrangian in the Einstein frame we write the five-dimensional metric as
\begin{align}
ds^2_5 &= e^{\frac{1}{\sqrt 3}\phi_1} ds^2_4 + e^{-\frac{2}{\sqrt 3}\phi_1}( dz+ A_1)^2, \label{eqn:metric5to4d} \\
A &= A_2 + \axone  dz. \label{pot5to4d}
\end{align}
From this
ansatz one finds (see for example \cite{Cremmer:1999du}) that the resulting four-dimensional Lagrangian takes the form
\begin{align}
\mathcal{L}_4 &= R_4 \star_4 \mathbf{1} - \frac{1}{2} \star_4 d \phi_1 \wedge  d\phi_1 - \frac{1}{2} e^{\frac{2}{\sqrt{3}} \phi_1} \star_4  d\chi_2 \wedge  d\chi_2  - \frac{1}{2} e^{-\sqrt{3}\phi_1}\star_4 F_1 \wedge F_1 \nn \\
&\quad\, \: - \: \frac{1}{2} e^{-\frac{1}{\sqrt{3}}\phi_1} \star_4 F_2 \wedge F_2 + \frac{1}{\sqrt{3}} \, \chi_2 \,  dA_2 \wedge  dA_2,
\label{lagrangian}
\end{align}
where
\begin{align}
\label{eqn:4dfieldstrengths}
F_1 &= d A_1, &
F_2 &= d A_2 - d\chi_2 \wedge A_1.
\end{align}
The scalars $\phi_1$ and $\chi_2$ parameterize an $\mathrm{SL}(2, \mathbb{R})/\mathrm{U}(1)$ coset.

One can easily reproduce the Lagrangian \eqref{lagrangian} from the $N=2$ prepotential formalism. To this end we start by recalling some basic facts about $N=2$ supergravity. The action of $N=2$ supergravity coupled to $n_v$ vector-multiplets is governed by a prepotential function $F$ depending on $(n_v + 1)$ complex scalars $X^I$ $(I=0,\,\ldots,\,n_v)$.
The bosonic degrees of freedom of $N=2$ supergravity are the metric $g_{\mu \nu}$, the complex scalars $X^I$ and a set of $(n_v+1)$ gauge fields $\check A^I_{\mu}$. The bosonic part of the action is given as \cite{Ceresole:1995jg}
\be
\mathcal{L}_4 =  R \star_4 \mathbf{1} - 2 g_{i\bar{j}} \star_4 dX^i \wedge  d\bar{X}^{\bar{j}} + \frac{1}{2} \check F^{I} \wedge \check G_{I},
\label{n=2}
\ee
where $\check F^I = d \check A^I$. The ranges of the indices are $i,j = 1, \ldots, n_v$, and  $ g_{i \bar{j}} = \partial_i \partial_{\bar{j}} K$ is the K\"ahler metric with the K\"ahler potential
\be
K = - \log \left[ -i (\bar{X}^I F_I - \bar{F}_I X^I) \right].
\ee
The two-forms $\check G_I$ are defined as
\be \check G_I = (\mbox{Re} N)_{IJ} \check F^J + (\mbox{Im} N)_{IJ}\star_4 \check F^J~, \ee where the complex symmetric matrix $N_{IJ}$ is constructed from the prepotential $F(X)$ as
\be N_{IJ} = \bar{F}_{IJ} + 2 i \frac{(\mbox{Im}F \cdot X)_I(\mbox{Im}F \cdot X)_J}{X \cdot\mbox{Im}F \cdot X}~,\ee
and $F_I = \partial_I F$ and $F_{IJ} = \partial_I \partial_J F.$  For the system we are interested in, the prepotential is
\be
F(X) = - \frac{(X^1)^3}{X^0}.
\ee
We fix the gauge $X^0 = 1$.  With the parameterization
\be
X^1= - \frac{1}{\sqrt{3}}{\chi _2}+i e^{-\frac{\phi_1 }{\sqrt{3}}},
\ee
the Lagrangian \eqref{n=2} takes the form
\begin{align}
\mathcal{L}_4 &= R_4 \star_4 \mathbf{1} - \frac{1}{2} \star_4 d \phi_1 \wedge  d\phi_1 - \frac{1}{2} e^{\tfrac{2}{\sqrt{3}} \phi_1} \star_4  d\chi_2 \wedge  d\chi_2  \nn \\ &\quad\,
- \: \frac{1}{2} \left( e^{-\sqrt{3} \phi_1} + e^{-\frac{\phi_1}{\sqrt{3}}} \chi_2^2\right)\check F^0 \wedge \star_4 \check F^0 \nn \\ &\quad\,
- \: \frac{3}{2} \: e^{-\frac{\phi_1}{\sqrt{3}}} \: \check F^1 \wedge \star_4\check F^1 -  \sqrt{3} \: e^{-\frac{\phi_1}{\sqrt{3}}} \: \chi_2 \: \check F^0 \wedge \star_4 \check F^1 \nn \\ &\quad\,
+ \:  \sqrt{3} \: \chi_2 \: \check F^1 \wedge  \check F^1~ +  \chi_2^2 \: \check F^0 \wedge  \check F^1 + \ \frac{1}{3\sqrt{3} }  \chi_2^3\: \check F^0 \wedge \check F^0.
\label{lag2}
\end{align}
With the field redefinition
\begin{align}
\check A^0&= A_1,   & \check A^1 &= \frac{1}{\sqrt{3}} \left(A_2 - \chi_2 A_1 \right),
\end{align}
the Lagrangian \eqref{lag2} becomes identical to \eqref{lagrangian}.

As noted above, the S$^3$ model has two dynamical vectors and two scalars. We now derive expressions for the four electromagnetic charges corresponding to the two vectors.
The equations of motion for the potentials $A_1$ and $A_2$ are
\begin{equation}\label{eqn:eom4A1}
d\big( e^{-\sqrt{3} \phi_1} \star_4 F_1)+e^{-\frac{1}{\sqrt{3}} \phi_1} \star_4 F_2\wedge d\chi_2 =0,
\end{equation}
and
\begin{equation}\label{eqn:eom4A2}
d\beta_2 \equiv d\big( e^{-\frac{1}{\sqrt{3}} \phi_1} \star_4 F_2 - \frac{2}{\sqrt{3}} \chi_2 d A_2 \big) = 0.
\end{equation}
Using \eqref{eqn:eom4A2}, one can rewrite \eqref{eqn:eom4A1} as the closure of a form $\beta_1$,
\begin{equation}\label{eqn:eom4A1-1}
d\beta_1 \equiv d\big( e^{-\sqrt{3} \phi_1} \star_4 F_1+e^{-\frac{1}{\sqrt{3}} \phi_1} \star_4 F_2 \chi_2 - \frac{1}{\sqrt{3}} d A_2 \chi_2^2\big)=0.
\end{equation}
We then use the closed forms $\beta_1$ and $\beta_2$ given by \eqref{eqn:eom4A1-1} and \eqref{eqn:eom4A2} to define conserved electric charges in asymptotically flat four-dimensional spacetimes as integrals over a two-sphere at spatial infinity $S^2_\infty$,
\begin{align}
Q_{1} &= \frac{1}{4 \pi}  \int_{S^2_{\infty}} \beta_1,&
Q_{2}  &= \frac{1}{4 \pi \sqrt{3}}  \int_{S^2_{\infty}} \beta_2.
\end{align}
In a similar fashion, from the Bianchi identities
\begin{align}
d F_1 &= 0,& d( F_2 + d \chi_2 \wedge A_1 ) &= 0,
\end{align}
for $A_1$ and $A_2$ we define the magnetic charges
\begin{align}
P_1 &=- \frac{1}{4 \pi}  \int_{S^2_{\infty}} F_1, &
\label{P1P2}
P_2 &=  \frac{1}{4 \pi \sqrt{3}}  \int_{S^2_{\infty}} F_2 + d\chi_2 \wedge A_1.
\end{align}
Note the minus sign in the definition of $P_1$. We work with sign conventions in which the static extremal black hole carrying positive $P_1$ and $Q_2$ charges
is BPS. From the M-theory point of view the electromagnetic charges correspond to the following brane charges\footnote{For more details on the brane interpretation see the brane intersection tables in \cite{Compere:2009zh}.}:
\begin{itemize}
\item $Q_1$ for Kaluza-Klein momentum (P) along the M-theory circle,
\item $P_2$ for Kaluza-Klein monopole charge (KKM) along the M-theory circle,
\item $Q_2$ for three equal M2 charges (M$2^3$),
\item $P_2$ for three equal M5 charges (M$5^3$).  
\end{itemize}
Thus, the $(P_1, Q_2)$ system with $P_1,\,Q_2 > 0$ corresponds to KKM -- M2$^3$, which is BPS. Similarly, the sign conventions for $Q_1$ and $P_2$ are chosen so that the extremal black hole carrying positive $Q_1$ and $P_2$ charges
is also BPS. The $(Q_1, P_2)$ system with $Q_1,\,P_2 > 0$ corresponds to
M5$^3$ -- P.

We now define two scalar charges for $\phi_1$ and $\chi_2$ as the radial derivatives of these fields at spatial infinity. As noted in section \ref{sec:NilpotentOrbits}, in order to make sure that charge matrices are in $\tilde{\mathfrak{p}}$ we must impose the condition that all scalars vanish at infinity. With this condition imposed, the scalar charges can simply be defined as
\begin{align}
\Sigma &= \lim_{r\to\infty}\frac{r\,\phi_{1}(r)}{\sqrt3},&
\Xi&= \lim_{r\to\infty}\frac{r\,\chi_{2}(r)}{\sqrt3}.
\end{align}

A consistent truncation of the S$^3$ model is obtained by setting
\begin{align}
\phi_1 &= \chi_2 = 0, & \star_4 F_1 &= \frac{1}{\sqrt{3}} dA_2,
\label{EMtruncation}
\end{align}
which reduces \eqref{lagrangian} to minimal $N=2, D=4$ supergravity, i.e.,  the pure Einstein-Maxwell theory. In section \ref{sec:EinsteinMaxwell}, we  discuss this consistent truncation  from the coset model point of view.

\subsection{Reduction on time}
\label{sec:reductionontime}
In this paper we exclusively work with stationary-axisymmetric spacetimes.  Therefore, we also assume the existence of a timelike Killing vector $\partial_t$ commuting with $\partial_z$. Now we can reduce the theory to three dimensions over this timelike Killing vector. The standard Kaluza-Klein ansatz for this reduction is
\begin{align}
\label{eqn:metric4dto3d}
ds^2_4  &= e^{\phi_2}ds^2_3  - e^{-\phi_2}(dt + \Atwo)^2, \\
\label{eqn:metricMaxwellto3d}
A_1 & =  B_1 + \Azero dt, \\
\label{eqn:Maxwellto3d}
A_2 & =  B_2+ \axtwo dt.
\end{align}
From this reduction we end up with three-dimensional Euclidean gravity coupled to five scalars and three one-forms. The one-forms $B_1, B_2,$ and $\omega_3$ can be dualized into
the scalars $\chi_5$, $\chi_4$, and $\chi_6$, respectively, in the notation of \cite{Compere:2009zh}. Upon dualization the Lagrangian in three dimensions becomes Euclidean gravity coupled to eight scalars. The scalars parameterize the pseudo-Riemannian coset $\mathrm{G}_{2(2)}/\tilde{K}$, with
\begin{align}
\tilde{K} = \mathrm{SO}_0(2,2) \cong  (\mathrm{SL}(2, \RR) \times \mathrm{SL}(2, \RR))/\mathbb{Z}_2.
\end{align}
For the derivation of the sigma model we follow the conventions of \cite{Compere:2009zh}. Relevant details are also presented in appendix \ref{sec:SigmaModel}.

Using the reduction ansatz (\ref{eqn:metric4dto3d}) we can calculate the mass and NUT charge explicitly in terms of the three-dimensional fields. Following \cite{Bossard:2008sw}, we define the Komar mass and NUT charge as
\begin{align}
M &= \frac{1}{8 \pi G_4} \int_{S^2_{\infty}} \star_4 K, & N &=  \frac{1}{8 \pi G_4} \int_{S^2_{\infty}}  K,
\end{align}
where $K=dg$, $g=g_{\mu\nu} \kappa^\mu dx^\nu$, and  $\kappa = \partial_t$. {From now on we restrict ourselves to flat three-dimensional base space with coordinates}
\be
{ds^2_3 = dr^2 + r^2 (d \theta^2 + \sin^2 \theta d\phi^2). \label{3dmetric} }
\ee
{From the metric ansatz (\ref{eqn:metric4dto3d}) and (\ref{3dmetric}) we find}
\begin{equation}
K = \partial_{\nu} g_{t \mu} dx^{\nu} \wedge dx^{\mu},
\end{equation}
which yields
\begin{equation}
M = -\frac{1}{8 \pi G_4} \int_{S^2_{\infty}} \partial_r e^{-\phi_2} \,\star_4 (dr \wedge dt),
\end{equation}
where, in our conventions, $\star_4 ( d r \wedge d t) = - e^{\phi_2}r^2 \sin(\theta)  d \theta \wedge d \phi$.
From asymptotic flatness we have
\begin{equation}
\phi_2 (r)= \frac{2 G_4 M}{r}+ \mathcal{O}\left(\frac{1}{r^2}\right).
\end{equation}
(For a detailed discussion of boundary conditions we refer the reader to \cite{Bossard:2008sw}.) From here on we simply set $2 G_4 = 1$, and thus,
\begin{align}
M = \lim_{r\to\infty}r\phi_{2}(r).
\end{align}
Calculation of the NUT charge proceeds in almost the same fashion.
Working out $N$ using the ansatz (\ref{eqn:metric4dto3d}) one finds \cite{Bossard:2008sw}
\begin{equation}
N =  -\frac{1}{4 \pi}  \int_{S^2_{\infty}} \partial_{\theta} (\omega_3)_{\phi}\, d \theta \wedge d \phi,
\end{equation}
where $(\omega_3)_{\phi}$ is the $\phi$-component of $\omega_3$. Demanding $N=0$, one sees that $ \partial_{\theta} (\omega_3)_{\phi}  \sim \mathcal{O}(1/r)$ at infinity.

To summarize, the ansatz
\begin{align}
ds^{2}_{5} &=
e^{\frac{1}{\sqrt3} \phi_{1}+\phi_{2}}ds^{2}_{3}
-e^{\frac{1}{\sqrt3} \phi_{1}-\phi_{2}}  (dt+\omega_3)^2
+e^{-\frac{2}{\sqrt3} \phi_{1}} ( dz +B_1 + \chi_1 dt)^{2},\label{eqn:canonicalform}
\\
\label{eqn:canonicalMaxwell}
A&=B_2+\chi_{3}dt + \chi_{2}dz,
\end{align}
describes stationary solutions of the S$^3$ model uplifted to five dimensions. It follows from four-dimensional asymptotic flatness that the electric and magnetic charges defined above can  also be expressed in terms of asymptotic values of the scalars. We find
\begin{align}
Q_1&=   \lim_{r \to \infty} \,r \chi_1(r),& 
Q_2&=    \lim_{r \to \infty} \, \frac{r \chi_3(r)}{\sqrt{3}},\nn\\
P_1&= \lim_{r \to \infty} \,r \,\chi_5(r),& 
P_2&= -\lim_{r \to \infty} \, \frac{r\, \chi_4(r)}{\sqrt{3}}. \label{P1P2Q1Q2}
\end{align}
Similarly, the NUT charge can also be expressed as \be
N = -\lim_{r\to\infty}r\chi_{6}(r).
\ee

\subsection{Five-dimensional lift and hyper-K\"ahler base space}
\label{sec:timereduction}
As reviewed above, the $N=2,\, D=4$ single modulus S$^3$ model arises as dimensional reduction of five-dimensional minimal supergravity. Equivalently all solutions of the S$^3$ model can be uplifted to five-dimensional minimal
supergravity. Supersymmetric solutions  of minimal five-dimensional supergravity have been completely classified \cite{Gauntlett:2002nw}. In the later sections we will discuss our findings in relation to the analysis of \cite{Gauntlett:2002nw}. To this end, we now take a small detour from the S$^3$ model and discuss supersymmetric solutions of minimal five-dimensional supergravity.  This section is a short summary of the results of \cite{Gauntlett:2002nw}. We refer the reader to this paper for more details.

The existence of a Killing spinor implies the existence of a timelike or null Killing vector.  In the case of a timelike Killing vector, five-dimensional solutions are most conveniently described as a timelike fibre over a four-dimensional Euclidean base space $\mathcal{B}$,
\begin{equation}
\label{eqn:metric5dto4dEuclidean}
ds_5^2 = -f^2 (dt + \omega)^2+f^{-1} ds_4^2(\mathcal{B}),
\end{equation}
where $f$ and $\omega$ are a function and a one-form on the base space $\mathcal{B}$, respectively. From the existence of the Killing spinor one can also infer the existence of three covariantly constant almost complex structures over the manifold $\mathcal{B}$ obeying the algebra of imaginary unit quaternions. The base space $\mathcal{B}$ is therefore in general a hyper-K\"ahler manifold.

Defining  $G^+$ and $G^-$ as the self-dual and anti-self-dual parts of the form $f d\omega$ with respect to the Euclidean metric on $\mathcal{B}$,
\be
f d\omega = G^+ + G^-,
\ee
the Maxwell field for supersymmetric spacetimes can be written as
\begin{equation}
\label{eqn:FieldStrength}
F = \sqrt{3} \,d\big(f (dt+\omega)\big)-\frac{2}{\sqrt{3}} G^+.
\end{equation}

It was shown in \cite{Gibbons:1987sp} that if a four-dimensional hyper-K\"ahler manifold admits a Killing vector that preserves the complex structures, then it must be a Gibbons-Hawking \cite{Gibbons:1979zt} metric
\begin{equation}
\label{eqn:GibbonsHawkingBaseSpace}
ds_4^2(\mathcal{B}) = H^{-1}\left( d z + \chi \right)^2 + H ds_3^2.
\end{equation}
The isometry direction that preserves the complex structures is $\partial_z$. The form of the metric is a $\mathrm{U(1)}$ fibration over a three-dimensional Euclidean flat space  $ds_3^2$. Here $\chi$ is a one-form on $\mathbb{R}^3$ and is determined from $H$ via
\begin{equation}
\label{eqn:GibbonsHawking}
\star_3 d H =  d \chi.
\end{equation}
This equation implies that $H$ is a harmonic function in three-dimensional Euclidean space.  Assuming that the Killing vector $\partial_z$ is a Killing vector of the full five-dimensional spacetime, the equations for $f$ and $\omega$ can be solved explicitly in terms of harmonic functions on $\mathbb{R}^3$ \cite{Gauntlett:2002nw}.
From now on we assume that $\partial_z$ is a Killing vector of the full five-dimensional spacetime and work exclusively with the Gibbons-Hawking form of the base space. Following \cite{Gauntlett:2002nw} we also write
\begin{equation}
\label{eqn:generalomega}
\omega = \omega_5 (d z+ \chi) + \omega_i dx^i ,
\end{equation}
where $\omega_5$ and $\omega_i $ are functions on $\mathbb{R}^3$.

The Gibbons-Hawking form thus naturally allows us to relate solutions of the four-dimensional S$^3$ model to five-dimensional minimal supergravity. It was conjectured in \cite{Gauntlett:2002nw} that the dimensional reduction of solutions of five-dimensional minimal supergravity with Gibbons-Hawking base gives the entire class of supersymmetric solutions of the S$^3$ model with the Killing spinor squaring to a timelike Killing vector.  To the best of our knowledge this conjecture has not yet been proven. Our analysis provides an interesting perspective on the conjecture---we show that all single center supersymmetric solutions of the S$^3$ model can be written in the Gibbons-Hawking form.
\subsection{Asymptotic frame}
As is previously discussed in the literature (see e.g., \cite{Elvang:2005sa}), solutions obtained in the formalism of \cite{Gauntlett:2002nw} generically turn out to have quite non-trivial asymptotic structure. In particular, the following situation arises. The five-dimensional metric
asymtotes to
\be\label{asymp-cross}
ds^2_{\rom{asymp}} = -\Big{(}dt + v_{\rom{H}} (dz + P \cos \theta d\phi) \Big{)}^2 + (dz + P \cos \theta d \phi)^2 + ds^2_{3}.
\ee
(This happens e.g., for the solution of \cite{Elvang:2005sa}.)
The cross term $v_{\rom{H}} \, (dz + P \cos \theta d\phi)\,dt$ arises from the fact that asymptotically $\omega_5 \rightarrow v_{\rom{H}}$. The presence of this term implies that the asymptotic frame is not at rest. In order to get the asymptotic frame at rest we need to do a coordinate transformation---a shift followed by a rescaling \cite{Elvang:2005sa}
\be
t = \gamma^{-1} \bar t~, \qquad z = \gamma (\bar z + v_{\rom{H}} \bar t)~, \qquad \gamma = \frac{1}{\sqrt{1-v_{\rom{H}}{}^2}}.\label{coordtrans1}
\ee
The asymptotic metric now takes the manifestly flat form
\be
ds^2_{\rom{asymp}} = \left( d\bar z + \frac{P}{\gamma} \cos \theta d\phi \right)^2 - d\bar t^2 + ds^2_3.
\ee

While using group theoretical methods we always work with manifestly asymptotically flat metrics. The above example illustrates the fact that in order to rewrite solutions obtained via the sigma-model in the Gibbons-Hawking form, certain coordinate transformations may be required. For solutions considered in this paper the linear shift
\be
z \rightarrow z + v t,\label{eqn:Zshift}
\ee
turns out to be sufficient. For each supersymmetric orbit below we will perform a coordinate transformation of the form \eqref{eqn:Zshift} to display our solutions manifestly in the Gibbons-Hawking form. This coordinate transformation has the effect of shifting $\chi_1$ by $v$,
\be
\chi_1 \rightarrow \chi_1 + v,
\ee
in the metric. The $t$-component of the Maxwell field will be shifted with the $z$-component. Comparing (\ref{eqn:canonicalform}) to (\ref{eqn:metric5dto4dEuclidean})  after taking into account the shift we find
\begin{align}
\label{eqn:fsquareshifted}
f^2&=e^{\frac{1}{\sqrt3} \phi_{1}-\phi_{2}}-e^{-\frac{2}{\sqrt3} \phi_{1}}(\chi_1 + v)^2,\\
\label{eqn:harmonicfuncshifted}
H&=f e^{\frac{1}{\sqrt3}\phi_1+\phi_2},\\
\omega_5&= - f^{-2} e^{-\frac{2}{\sqrt3}\phi_1}(\chi_1 + v),\\
\label{eqn:chi}
\chi&= B_1.
\end{align}
To obtain these expression we have set  $\omega_3 =0$ to eliminate the NUT charge. The condition \eqref{eqn:GibbonsHawking} yields
\begin{align}
d\Big(f\,e^{\frac{1}{\sqrt3} \phi_1+\phi_2} \Big)= \star_3
dB_1
.
\end{align}
Comparing the Maxwell fields  gives
\begin{align}
\label{eqn:chi2}
d \chi_2 &= \sqrt{3}\, d(f \omega_5 )-\frac{1}{\sqrt{3}}f \,d\omega_5+ \frac{1}{\sqrt3}f\, \omega_5\, H^{-1} \,dH,\\¬†
\label{eqn:chi3}
d\chi_3 + v d\chi_2 &= \sqrt{3}\, df,\\¬†
\label{eqn:B2}
d B_2&= d\chi_2 \wedge \chi+ \frac{2}{\sqrt{3}} f\omega_5  d \chi+ \frac{1}{\sqrt{3}} f H \star_3 d\omega_5.
\end{align}
Checking equations \eqref{eqn:chi2}--\eqref{eqn:B2} for  supersymmetric solutions  provides a non-trivial test on the consistency of our approach.


\setcounter{equation}{0}
\section{Orbit structure of G$_{2(2)}$}
In this section we introduce $\mf g_{2(2)}$, the Lie algebra of the hidden symmetry group G$_{2(2)}$. We discuss the reductive decomposition following from the coset structure G$_{2(2)}/\mathrm{SO}_0(2,2)$ in section \ref{genG2} and the associated nilpotent orbits in sections \ref{sec:orbitstructure}, \ref{sec:orbitstructure1}, and \ref{sec:orbitstructure2}.

\subsection{Generalities on G$_{2(2)}$ }
\label{genG2}
The Lie algebra $\mathfrak{g}_{2(2)}$ is the
split real form of the complex Lie algebra $\mathfrak{g}_{2}$. With rank 2 and dimension 14 it is the smallest of the exceptional Lie algebras.
It is generated by two triples of Chevalley generators,
\begin{align} \label{g2basdef0}
&(h_1,\,e_1,\,f_1), & &(h_2,\,e_2,\,f_2),
\end{align}
satisfying the Chevalley relations
\begin{align} \label{chevrel}
[h_1,\,e_1]&=2e_1, & [h_2,\,e_1]&=-3e_1, &
[h_1,\,f_1]&=-2f_1, & [h_2,\,f_1]&=3f_1, \nn\\
[h_1,\,e_2]&=-e_2, & [h_2,\,e_2]&=2e_2, &
[h_1,\,f_2]&=f_2, & [h_2,\,f_2]&=-2f_2, \nn\\
[e_1,\,f_1]&=h_1, & [e_1,\,f_2]&=0, &
[e_2,\,f_2]&=h_2, & [e_2,\,f_1]&=0.
\end{align}
The elements $h_1$ and $h_2$ span the Cartan subalgebra $\mf{h} \subset \mathfrak{g}_2$.
We define the additional basis elements by
\begin{align} \label{g2basdef}
e_3&=[e_1,\,e_2], & e_4&=
[e_3,\,e_2], & e_5&=
[e_4,\,e_2], & e_6&=
[e_1,\,e_5],\nn\\
f_3&=[f_2,\,f_1], & f_4&=
[f_2,\,f_3], & f_5&=
[f_2,\,f_4], & f_6&=
[f_5,\,f_1].
\end{align}
The  elements $e_1,\,\ldots,\,e_6$ ($f_1,\,\ldots,\,f_6$)
are the root vectors associated to the positive (negative) roots
$\pm\alpha_1,\,\ldots,\,\pm\alpha_6$.
The complex span of the 14 basis elements gives the complex Lie algebra $\g_2$, whereas the real span gives the split real form $\g_{2(2)}$.
We thus have the  triangular decomposition
\begin{align}
\mf{g}_{2(2)} = {\mf{m}} \oplus \mf{h} \oplus {\mf{n}}, \label{decomp2}
\end{align}
as a direct sum of subspaces (but not a direct sum of subalgebras),
where ${\mf{m}}$ and ${\mf{n}}$ are spanned by root vectors associated to positive and negative roots, respectively. All the commutation relations for $\mathfrak{g}_{2(2)}$ can be derived from the Chevalley relations (\ref{chevrel}), the definitions (\ref{g2basdef}), and the Serre relations
\begin{align} \label{serrel}
[e_1,\,e_3]=[e_5,\,e_2]=[f_1,\,f_3]=[f_5,\,f_2]=0.
\end{align}

Since we are interested in compactifying five-dimensional minimal supergravity to three dimensions over one spacelike and one timelike Killing direction, we consider the involution $\tau$ of $\mf{g}_{2(2)}$ given by
\begin{align}\label{invol1}
\tau(e_1)&=f_1, & \tau(e_2)&=-f_2, \nn\\
\tau(h_1)&=-h_1, & \tau(h_2)&=-h_2.
\end{align}
When this involution is integrated to the group, it has the subgroup $\mathrm{SO}_0(2,2)$ as the fixed point set. Accordingly, the involution $\tau$ defines the
coset G$_{2(2)}/\mathrm{SO}_0(2,2)$.
It follows that
\begin{align}\label{invol2}
\tau(e_3)&=f_3, & \tau(e_4)&=f_4, &
\tau(e_5)&=f_5, & \tau(e_6)&=-f_6,
\end{align}
and thus the involution $\tau$ differs from the Chevalley involution (which would be relevant for compactifying over two spacelike directions) by the signs of $f_1,\,f_3,\,f_4,\,f_5$.
The action of $\tau$ on the negative root vectors follows from the property that an involution squares to the identity map.

Let $\tilde{\mf{k}}$ denote the subalgebra of $\mf{g}_{2(2)}$ pointwise fixed by the involution $\tau$, which is the Lie algebra of $\tilde{K}$, and thus $\tilde{\mf{k}}=\mf{sl}(2,\RR)\oplus\mf{sl}(2,\RR).$
As a basis of $\tilde{\mf{k}}$, we define the linear combinations
\begin{align}
k_1 &= e_1+f_1, & k_2&=e_2-f_2, & k_3&=e_3+f_3,\nn\\
k_4 &= e_4+f_4, & k_5&=e_5+f_5, & k_6&=e_6-f_6 .
\end{align}
The orthogonal complement $\tilde{\mf{p}}$ of $\tilde{{\mf{k}}}$ in $\mathfrak{g}_{2(2)}$ with respect to the Killing form is defined as the eigenspace of $\tau$ with eigenvalue $-1$. It is spanned by $h_1,\,h_2$ and
the linear combinations
\begin{align}
p_1 &= e_1-f_1, & p_2&=e_2+f_2, & p_3&=e_3-f_3,\nn\\
p_4 &= e_4-f_4, & p_5&=e_5-f_5, & p_6&=e_6+f_6.
\end{align}
Beside the triangular decomposition (\ref{decomp2})
we thus also have the reductive decomposition
\begin{align}
\mf{g}_{2(2)} = \tilde{\mf{k}} \oplus \tilde{\mf{p}}.  \label{decomp1}
\end{align}
As in (\ref{decomp2}), this is not a direct sum of subalgebras, but of subspaces that do not commute with each other. We have
\begin{align} \label{reduktivt}
[\tilde{\mf{k}},\,\tilde{\mf{k}}]&=\tilde{\mf{k}},&
[\tilde{\mf{k}},\,\tilde{\mf{p}}]&=\tilde{\mf{p}},&
[\tilde{\mf{p}},\,\tilde{\mf{p}}]&=\tilde{\mf{k}}.
\end{align}

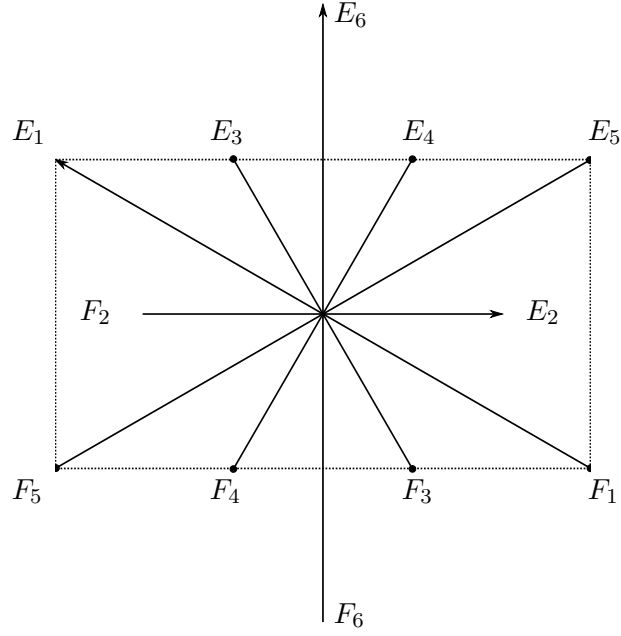
\begin{figure}[h]
\begin{center}
\begin{overpic}[scale=0.8]{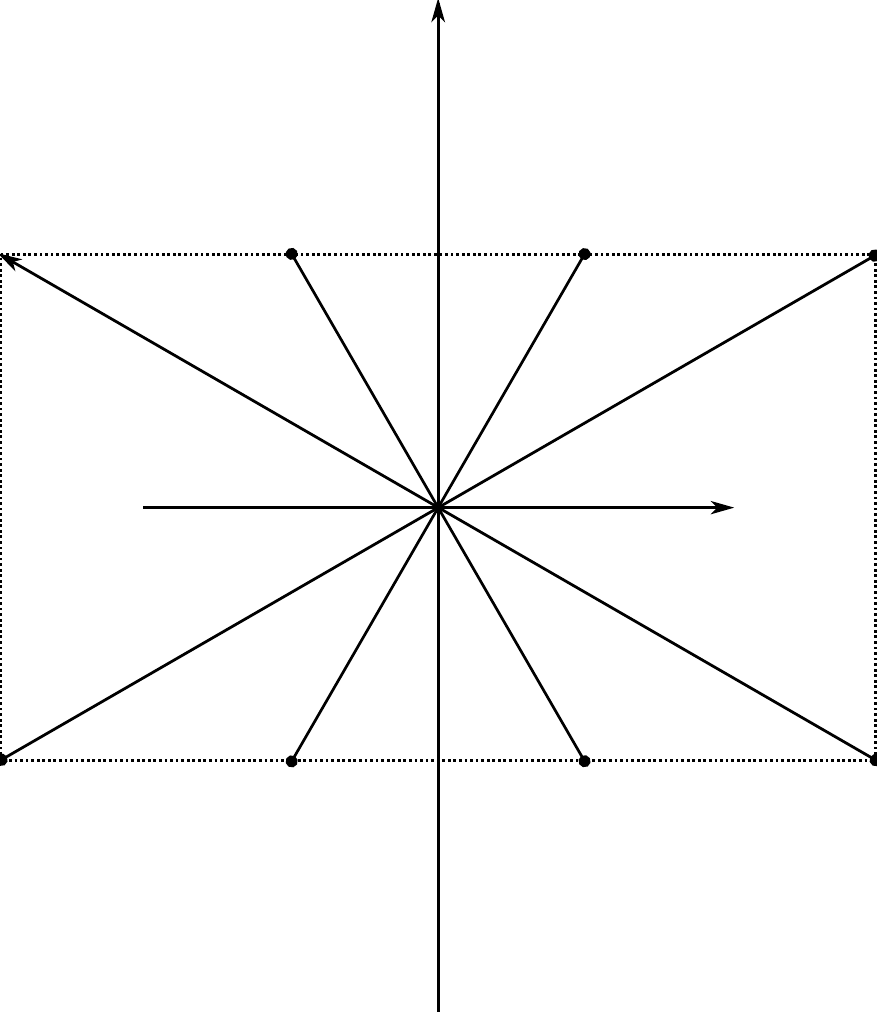}
\put(76,49){$E_2$}
\put(4,49){$F_2$}
\put(45,97){$E_6$}
\put(86,78){$E_5$}
\put(56,78){$E_4$}
\put(25,78){$E_3$}
\put(-7,78){$E_1$}
\put(86,20){$F_1$}
\put(56,20){$F_3$}
\put(25,20){$F_4$}
\put(-7,20){$F_5$}
\put(45,0){$F_6$}

\end{overpic}
\caption{\small
The roots of $\mathfrak{g}_{2}$ given as vectors in the two-dimensional root space, dual to the Cartan subalgebra $\mathfrak{h}$. The positive (negative) roots
$\pm\alpha_1,\,\ldots,\,\pm\alpha_6$ correspond to the root vectors $e_1,\,\ldots,\,e_6$ ($f_1,\,\ldots,\,f_6$),
which
under the automorphism $\varphi$ are mapped into
$E_1,\,\ldots,\,E_6$ ($F_1,\,\ldots,\,F_6$). Applying this automorphism, we can associate the horizontal and vertical axes with the subalgebra $\tilde{\mf{k}}$, and the rectangle with the representation space
$\tilde{\mf{p}}$.}
\label{fig:g2rootdiagram}
\end{center}
\end{figure}

When we say that an element $x\in\g_{2(2)}$ is nilpotent we always refer to the adjoint action of $x$ on the whole of $\g_{2(2)}$. Thus it means that there is an integer $n$ such that
\begin{align}
(\text{ad }x)^n(y)=0
\end{align}
for all $y\in\g_{2(2)}$.
It follows from (\ref{g2basdef}) and (\ref{serrel}) that any element in
${\mf{m}}$ or ${\mf{n}}$ is nilpotent, which makes the triangular decomposition (\ref{decomp2}) useful for studying nilpotent elements. On the other hand,
we are interested in nilpotent elements in the subspace $\tilde{\mf{p}}$, which is given by the reductive decomposition (\ref{decomp1}).
Therefore we introduce the automorphism
\begin{align}
\varphi&=\text{Ad }\big(\tfrac{\pi}{8\sqrt{2}}(-6p_1+2p_3-p_4+p_5)+\tfrac{\pi}{16}(6k_2+k_6)\big),
\end{align}
of $\mf{g}_{2(2)}$ that partially maps ${\mf{m}}$ and
${\mf{n}}$ into $\tilde{\mf{p}}$, and also the Cartan subalgebra $\mf{h}$ into $\tilde{\mathfrak{k}}$.
It acts as
\begin{align} \label{g2auto2}
h_1 \mapsto H_1 &\equiv \tfrac16 k_5,&
h_2 \mapsto H_2 &\equiv -\tfrac12 k_3 -\tfrac1{4}k_5,\nn\\
e_1 \mapsto E_1 &\equiv -\tfrac1{12} p_5 +\tfrac1{2}h_1+\tfrac12 h_2,&
f_1 \mapsto F_1 &\equiv \tfrac1{12} p_5 +\tfrac1{2}h_1+\tfrac12 h_2,\nn\\
e_2 \mapsto E_2 &\equiv \tfrac34 k_1+\tfrac14 k_2+\tfrac18 k_4+\tfrac18 k_6,&
f_2 \mapsto F_2 &\equiv \tfrac34 k_1-\tfrac14 k_2+\tfrac18 k_4-\tfrac18 k_6,\nn\\
e_3 \mapsto E_3 &\equiv -\tfrac34 p_1 +\tfrac1{4}p_2+\tfrac18 p_4+\tfrac18 p_6,&
f_3 \mapsto F_3 &\equiv \tfrac34 p_1 +\tfrac1{4}p_2-\tfrac18 p_4+\tfrac18 p_6,\nn\\
e_4 \mapsto E_4 &\equiv -p_3-3h_1-h_2,&
f_4 \mapsto F_4 &\equiv p_3-3h_1-h_2,\nn\\
e_5 \mapsto E_5 &\equiv -\tfrac32 p_1+\tfrac32p_2-\tfrac34p_4-\tfrac14p_6,&
f_5 \mapsto F_5 &\equiv \tfrac32 p_1+\tfrac32p_2+\tfrac34p_4-\tfrac14p_6,\nn\\
e_6 \mapsto E_6 &\equiv \tfrac32 k_1+\tfrac32 k_2-\tfrac34 k_4-\tfrac14 k_6,&
f_6 \mapsto F_6 &\equiv \tfrac32 k_1-\tfrac32 k_2-\tfrac34 k_4+\tfrac14 k_6.
\end{align}
Beside the Cartan subalgebra $\mathfrak{h}$, also the root vectors $e_2,\,f_2,\,e_6,\,f_6$ are mapped into $\tilde{\mf{k}}$. The corresponding roots $\pm\alpha_2$ and $\pm\alpha_6$ are vectors along the horizontal and vertical axes in the root diagram in Figure \ref{fig:g2rootdiagram}. Since $\pm \alpha_2$ are orthogonal to $\pm \alpha_6$, the
$\sl(2,\,\mathbb{R})$ subalgebras spanned by $(e_2,\,f_2,\,h_2)$ and
$(e_6,\,f_6,\,h_6)$ commute with each other.

It follows from (\ref{reduktivt}) that
the adjoint action
of $\tilde{\mathfrak{k}}$ on
$\tilde{\mf{p}}$
is an irreducible representation of $\tilde{\mathfrak{k}}$.
Knowing that $\tilde{\mathfrak{k}}$ is isomorphic to
$\mf{sl}(2,\mathbb{R}) \oplus \mf{sl}(2,\mathbb{R})$, we
may thus use the representation theory of $\mathfrak{sl}(2, \mathbb{R})$ to describe $ \tilde{\mathfrak{p}}$ under the adjoint action
of $\tilde{\mathfrak{k}}$.
In fact, $E_5$
is the highest weight of ${\tilde{\mf{k}}}$. By acting with $F_2$ and $F_6$ we generate the representation $(\bf 4,2)$, indicated by a rectangle in Figure \ref{fig:g2rootdiagram}.

The complex Lie group $\mathrm{G}_{2}$ is the automorphism group of the Lie algebra $\g_2$.
It is easy to see that for any $x \in \g_2$ and any automorphism
$g \in \rm{G}_2$ we have
\begin{align}
\text{ad }g(x) = g \circ (\text{ad }x)\circ g^{-1}.
\end{align}
Thus, if we consider the automorphism $g$ as an element
in $\mathrm{GL}(14, \mathbb{C})$, and $x$ as an element in $\gl(14, \mathbb{C})$, we can write the
action of $g$ on $x$ as
\begin{align} \label{konjugering}
x \mapsto gxg^{-1}.
\end{align}
We say that $g$ acts on $x$ by {\it conjugation}, and when $g=\exp a$ for some
$a \in \g_2 \subset \gl(14, \mathbb{C})$, we denote the map (\ref{konjugering}) by $\text{Ad }a$.

A nilpotent $\mathrm{G}_{2}$-orbit is defined as the set
\begin{align}
\mathcal{O}=\{ gxg^{-1}\,|\, g \in \mathrm{G}_{2}\}
\end{align}
for some nilpotent element $x \in \g_2$, which is then a {\it representative} of the orbit $\mathcal{O}$. By restricting
$g$ to $\mathrm{G}_{2(2)}$ and $x \in \g_{2(2)}$, the $\mathrm{G}_{2}$-orbit gives rise to at least
one $\mathrm{G}_{2(2)}$-orbit, and by further restricting
$g$ to $\tilde{K}$, each $\mathrm{G}_{2(2)}$-orbit may split into different
$\tilde{K}$-orbits.
We describe this in more detail in the following subsections.

\subsection{$\mathrm{G}_{2}$-orbits}
\label{sec:orbitstructure}
Let $\g$ be an arbitrary complex semisimple Lie algebra with Cartan subalgebra $\mathfrak{h}$.
For each nilpotent orbit $\mathcal{O}$ there is a triple $(e,\,f,\,h)$ of elements, where $e \in \mathcal{O}$, such that
\begin{align}
[h,\,e]&=2e, & [h,\,f]&=-2f, & [e,\,f]&=h.
\end{align}
Following \cite{Collingwood:1993} we call this a \textit{standard triple}.
We can always take $h$ to be in the Cartan subalgebra $\mf{h}$.
Then $h$ is characterized by the eigenvalues $\alpha_i(h)$ of the adjoint action of $h$ on the simple root vectors,
\begin{align}
[h,\,e_i]&=\alpha_i(h)e_i, & i=1,\,2,\,\ldots,\,\text{rank }\g.
\end{align}
Furthermore, we can always find a standard triple such that $\alpha_i(h) \in \{0,\,1,\,2\}$. The eigenvalues $\alpha_i(h)$ then determines the nilpotent orbit uniquely. For $\g_2$ there are only two such eigenvalues, and we will refer to the pair $(\alpha_1(h),\,\alpha_2(h))$ as the  {\it$\alpha$-label} of the nilpotent $\mathrm{G}_2$-orbit.

There are four (nonzero) nilpotent $\mathrm{G}_2$-orbits, with $\alpha$-labels $(1,\,0)$, $(0,\,1)$, $(2,\,0)$ and $(2,\,2)$. From the $\alpha$-labels $(1,\,0)$ and $(0,\,1)$ we can easily construct corresponding standard triples $(e,\,f,\,h)$ where $e=m e_6$ and $e=m e_4$, respectively,
for an arbitrary nonzero complex number $m$.
To get representatives that are in $\tilde{\mf{p}}$, we
first note that (up to normalization) $e_6$ is conjugate to $e_1$ by a Weyl reflection,
\begin{align}
\text{Ad}(-\tfrac{\pi}4 p_5)(\tfrac16 e_6)=e_1.
\end{align}
Thus we can choose $me_6$ as well as $me_1$ as representative of the orbit with $\alpha$-label $(1,\,0)$.
Then applying the automorphism $\varphi$ to $me_1$ and $me_4$ gives $mE_1$ and $mE_4$, which are elements in $\tilde{\mf{p}}$.

By applying Weyl reflections we can map any long root to any other long root and any short root to any other short root. This means that the two orbits with $\alpha$-labels $(1,\,0)$ and $(0,\,1)$ already contain all root vectors in $\g_2$. To get representatives of the remaining two orbits, we need to take linear combinations of root vectors associated with different roots. It suffices to consider linear combinations of only two positive root vectors, with arbitrary nonzero complex coefficients $m$ and $n$. For the orbit with $\alpha$-label $(2,\,2)$
we then get $me_1+ne_2$ as a representative, whereas for the orbit with $\alpha$-label $(2,\,0)$ we get three possibilities,
\begin{align} \label{20representanter}
&me_1+ne_4,&&me_3+ne_5,&&me_1+ne_5.
\end{align}
Since these three elements belong to the same orbit they must be conjugate to one another. The first two are related by a Weyl reflection
\begin{align}
\text{Ad}(\tfrac{\pi}2 k_2)(m e_3+ne_5)=\tfrac12 m e_4 +6n e_1.
\end{align}
To relate the third expression in (\ref{20representanter}) to one of the first two, we can take for example
\begin{align}
\text{Ad} (\pm\tfrac{3\pi}{4} k_2) (e_3 + \tfrac16e_5)&= \sqrt{2}(\pm e_1 - \tfrac16e_5).
\end{align}

\subsection{$\mathrm{G}_{2(2)}$-orbits}
\label{sec:orbitstructure1}
In the preceding subsection we discussed  the complex Lie algebra $\mathfrak{g}_2$. Now we turn to  the algebra we are interested in, $\g_{2(2)}$, the split real form of $\g_2$. Given two  nilpotent elements $x$ and $y$ in $\g_{2(2)}$, it may happen that $g x g^{-1}=y$ for some $g \in \mathrm{G}_{2}$, but not for any $g \in \mathrm{G}_{2(2)}$.
This does not happen for the $\mathrm{G}_2$-orbits with $\alpha$-labels $(1,\,0)$, $(0,\,1)$, and $(2,\,2)$, so each of them contains just one single $\mathrm{G}_{2(2)}$-orbit. We call them $\mathcal{O}_1$, $\mathcal{O}_2$, and
$\mathcal{O}_5$, respectively.
On the other hand, the $\mathrm{G}_2$-orbit with $\alpha$-label $(2,\,0)$ splits into two different $\mathrm{G}_{2(2)}$-orbits. As we will see next, the two orbits are distinguished by relative signs of the coefficients $m$ and $n$ in the expression $m e_1+n e_4$.

Consider the automorphism
\begin{align}
\chi=\text{Ad }\tfrac{i\pi}{2}(2h_1+h_2)
\end{align}
of $\g_2$, which maps $e_1$ to $ie_1$ and $f_1$ to $-if_1$, leaving $e_2$ and $f_2$ invariant. It follows that
\begin{align}
p_n &\mapsto ik_n, & k_n &\mapsto ip_n, & \text{for }n&=1,\,3,\,4,\,5,\nn\\
p_n &\mapsto p_n, & k_n &\mapsto k_n,&  \text{for }n&=2,\,6,
\end{align}
under this automorphism.
Then $\mathfrak{k}=\chi(\tilde{\mathfrak{k}})=\su(2)\oplus\su(2)$ is the maximal compact subalgebra of $\g_{2(2)}$, and
$\mathfrak{p}=\chi(\tilde{\mathfrak{p}})$ the orthogonal complement with respect to the Killing form.
Let $\mathfrak{k}_\mathbb{C}=\mathfrak{a}_1\oplus\mathfrak{a}_1$ and $\mathfrak{p}_\mathbb{C}$ be the complexifications of these subspaces, and let
$K_\mathbb{C}$ be the complexification of the maximal compact subgroup.
Now
the nilpotent $\mathrm{G}_{2(2)}$-orbits are in one-to-one correspondence with the
nilpotent $K_\mathbb{C}$-orbits in $\mathfrak{p}_\mathbb{C}$, by the so called
Cayley transform \cite{Collingwood:1993}.
We will use this one-to-one correspondence
to find representatives of the two different
$\mathrm{G}_{2(2)}$-orbits with $\alpha$-label $(2,\,0)$.

To define the Cayley transform we need to consider standard triples $(e,\,f,\,h)$ such that
\begin{align}
\theta(h)&=-h, & \theta(e)&=-f, & \theta(f)&=-e,
\end{align}
where $\theta$ is the Chevalley involution.
Such standard triples are called {\it Cayley triples}. The Cayley transform of a Cayley triple $(e,\,f,\,h)$ is now defined as the standard triple $(e',\,f',\,h')$ where
\begin{align}
e'&=\tfrac12(e+f+ih),&
h'&=i(e-f),&
f'&=\tfrac12(e+f-ih).
\end{align}
It follows that $h' \in \mf{k}_{\mathbb{C}}$ and $e',\,f'\in  \mf{p}_{\mathbb{C}}$.

We can define a Cartan subalgebra of ${\mathfrak{k}_{\mathbb{C}}}$ as a subspace of
$(\chi \circ \varphi)(\mathfrak{h})$ such that the
simple roots of ${\mathfrak{k}_{\mathbb{C}}}=\mathfrak{a}_1\oplus\mathfrak{a}_1$ correspond to the roots $\alpha_6$ and $\alpha_2$ of $\g_2$.
Furthermore, we can take $h'$ to be in this Cartan subalgebra of
$\mf{k}_{\mathbb{C}}$
and conjugate
the standard triple
$(e',\,f',\,h')$
so that all eigenvalues
$\alpha_6(h'),\,\alpha_2(h')$
belong to the set
\{0,\,1,\,2,\,3,\,4,\,8\}
for all orbits.
Then the pair $(\alpha_6(h'),\,\alpha_2(h'))$ determines the ${K}_{\mathbb{C}}$-orbit uniquely, as well as the corresponding $\mathrm{G}_{2(2)}$-orbit, and we will refer to this pair as the $\beta${\it -label} of the
$\mathrm{G}_{2(2)}$-orbit.
The $\beta$-labels for the nilpotent $\mathrm{G}_{2(2)}$-orbits have been computed in \cite{Djokovic},
and we present them in Table \ref{g22orbitar}.
As we have already mentioned, there are two different $\mathrm{G}_{2(2)}$-orbits with
the same  $\alpha$-label $(2,\,0)$. One of them, which we call $\mathcal{O}_3$, has $\beta$-label $(0,\,4)$ and the other one, which we call
$\mathcal{O}_4$, has $\beta$-label $(2,\,2)$.
From the $\beta$-labels we can construct the corresponding standard triples
$(e',\,f',\,h')$.
For $\mathcal{O}_3$ we take
the standard triple
\setlength\arraycolsep{1.4pt}
\begin{align}
e'&=-\tfrac12 i k_3-\tfrac1{12}ik_5-h_1,&
h'&=-ip_3+\tfrac1{6}ip_5,&
f'&=\tfrac12 i k_3+\tfrac1{12}ik_5-h_1,
\end{align}
with the inverse
Cayley transform
\begin{align}
e&=\tfrac12(e'+f'-ih')&
h&=-i(e'-f')&
f&=\tfrac12(e'+f'+ih')\nn\\
&=-\tfrac12p_3+\tfrac1{12}p_5-h_1&
&=-k_3-\tfrac16 k_5&
&=\tfrac12p_3-\tfrac1{12}p_5-h_1\nn\\
&=F_1+\tfrac12 E_4,&
&=2H_1+2H_2,&
&=E_1+\tfrac12 F_4,
\end{align}
and for $\mathcal{O}_4$ we take the standard triple
\begin{align}
e'&=-\tfrac12 i k_3+\tfrac1{12}ik_5-h_1,&
h'&=-i p_3-\tfrac1{6}ip_5,&
f'&=\tfrac12 i k_3-\tfrac1{12}ik_5-h_1,
\end{align}
with the inverse Cayley transform
\begin{align}
e&=\tfrac12(e'+f'-ih')&
h&=-i(e'-f')&
f&=\tfrac12(e'+f'+ih')\nn\\
&=-\tfrac12p_3-\tfrac1{12}p_5-h_1&
&=-k_3+\tfrac16 k_5&
&=\tfrac12p_3+\tfrac1{12}p_5-h_1\nn\\
&=E_1+\tfrac12 E_4,&
&=4H_1+2H_2,&
&=F_1+\tfrac12 F_4.
\end{align}
Acting with $\text{Ad}(\tfrac{\pi}2 (E_1-F_1))$ we get
\begin{align}
\text{Ad}(\tfrac{\pi}2 (E_1-F_1))(F_1+\tfrac12 E_4)&=-E_1+\tfrac12 E_4,
\end{align}
and we see that it is indeed the relative sign of the coefficients of $E_1$ and $E_4$ that distinguishes between these two $\mathrm{G}_{2(2)}$-orbits.

To summarize, there are five nonzero $\mathrm{G}_{2(2)}$-orbits,
$\mathcal{O}_1,\,\ldots,\,\mathcal{O}_5$.
In Table \ref{g22orbitar} we list for each orbit the $\alpha$-label,
the $\beta$-label and a suitable representative in $\tilde{\mathfrak p}$.
We also list the dimensions of the $\mathrm{G}_{2(2)}$-orbits as well as of the corresponding $\tilde{K}$-orbits (which we will say more about in the following subsection).

\setlength{\arraycolsep}{4pt}

{\renewcommand{\arraystretch}{1.5}

\begin{table}[h!]
\begin{center}
\begin{equation}
\begin{array}{c|c|c|c|c|c}
\text{$\mathrm{G}_{2(2)}$-orbit} & 
\text{$\alpha$-labels}
& 
\text{$\beta$-labels}
&\text{representative $x$}&\text{dim} ( \mathrm{G}_{2(2)} \cdot x) &\text{dim}(\tilde K \cdot x)
\nn\\ \hline
\mathcal{O}_1 & (1,\,0)  &(1,\,1)  &E_1 & 6& 3\nn\\ \hline
\mathcal{O}_2 & (0,\,1)  &(1,\,3)  &E_4 & 8& 4\nn\\ \hline
\mathcal{O}_3 & (2,\,0)  &(2,\,2)  &E_4-E_1  &10& 5\nn\\
\mathcal{O}_4 & (2,\,0)  &(0,\,4)  &E_4+E_1  &10& 5\nn\\ \hline
\mathcal{O}_5 & (2,\,2) & (4,\,8) & E_1+E_2  &12& 6
\end{array}
\end{equation}
\caption{\small The five nonzero $\mathrm{G}_{2(2)}$-orbits.}
\label{g22orbitar}
\end{center}
\end{table}

}

Regarding the fifth orbit $\mathcal{O}_5$, any element $x$ in this orbit obeys
\begin{equation}
x^7=0,
\end{equation}
and hence has a too high nilpotency degree to be generated as the extremal limit of some non-extremal regular black hole. We henceforth ignore this orbit.\footnote{As discussed in \cite{Berkooz:2008rj}, $\mathcal{O}_5$ contains
the supersymmetric
G\"odel black hole solution of
\cite{Gauntlett:2002nw}.}

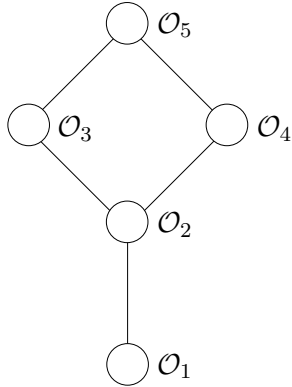
\begin{figure}[h!]
\begin{center}
\begin{overpic}[scale=0.4]{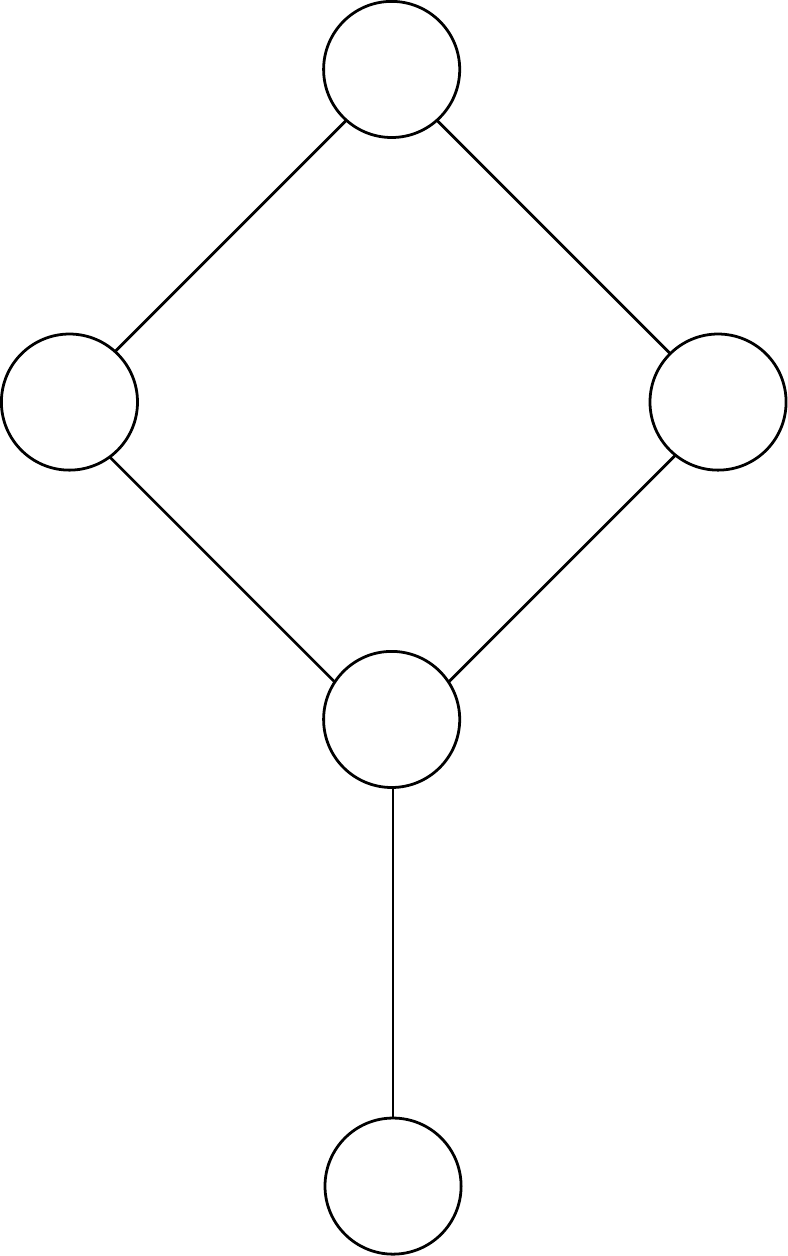}
\put(39,3){$\mathcal{O}_1$}
\put(39,40){$\mathcal{O}_2$}
\put(39,92){$\mathcal{O}_5$}
\put(65,65){$\mathcal{O}_4$}
\put(13,65){$\mathcal{O}_3$}
\end{overpic}
\caption{\small Hasse diagram for the partial ordering of the nilpotent orbits in $\mathfrak{g}_{2(2)}$.}
\label{fig:partialordering}
\end{center}
\end{figure}

One can impose a partial ordering on the different real nilpotent orbits. In \cite{Bossard:2009at} this is formulated in terms of a stratification on the space of black hole solutions. Indeed, all smaller orbits  lie in the closure of a bigger orbit and for $\mathfrak{g}_{2(2)}$ this is illustrated in Figure \ref{fig:partialordering} in the form of a Hasse diagram. For example, the solutions in $\mathcal{O}_1$ can be constructed as limits of the solutions in $\mathcal{O}_2$.

\subsection{$\tilde{K}$-orbits}
\label{sec:orbitstructure2}

By restricting the $\mathrm{G}_{2(2)}$ action to its subgroup $\tilde{K}$, the $\mathrm{G}_{2(2)}$-orbits $\mathcal{O}_3$ and $\mathcal{O}_4$ split into two
$\tilde{K}$-orbits each. The $\mathrm{G}_{2(2)}$-orbit $\mathcal{O}_3$ splits into two orbits that we call
$\mathcal{O}_{3K}$ and $\mathcal{O}_{3K}'$ with representatives
$E_4-E_1$ and $F_4+E_1$, respectively. These must be in different
$\tilde{K}$-orbits since $E_4-E_1$ commutes with
$E_6$, whereas $F_4+E_1$ commutes with $F_6 - F_2$, and as we have seen, $E_6$ is in a different $\mathrm{G}_{2(2)}$-orbit than $F_6 - F_2$.
Similarly, $\mathcal{O}_4$ splits into two different $\tilde{K}$-orbits that we call $\mathcal{O}_{4K}$ and $\mathcal{O}_{4K}'$ with representatives
$E_4+E_1$ and $F_4-E_1$, respectively. Again, one can find a standard triple $(e,\,f,\,h)$ for each of these four
$\tilde{K}$-orbits, such that $e$ is an element of the orbit. Within the $\mathrm{G}_{2(2)}$-orbit, each 
$\tilde{K}$-orbit is then characterized by the pair of eigenvalues $(\alpha_6(h),\,\alpha_2(h))$, which we will refer to as the $\gamma${\it-label} of the orbit. It turns out that the $\gamma${-labels} that distinguish between the two different $\tilde{K}$-orbits within each of $\mathcal{O}_3$ and $\mathcal{O}_4$
are $(0,\,4)$ and $(2,\,2)$, the same as the $\beta${-labels} that distinguish between
$\mathcal{O}_3$ and $\mathcal{O}_4$. This is illustrated in Table \ref{korbitar}.

As we will see, physical solutions in $\mathcal{O}_3$ and $\mathcal{O}_4$ all belong to the
$\mathcal{O}_{3K}$ and $\mathcal{O}_{4K}'$ orbits for which the $\beta$-label and the
$\gamma$-label coincide. This seems to be a generic property of extremal black holes that was observed in \cite{Bossard:2009we}.

\begin{table}[h!]
\begin{center}
\setlength\arraycolsep{10pt}
\begin{equation}
\renewcommand{\arraystretch}{1.5}
\begin{array}{c|c|c|c}
&
{\renewcommand{\arraystretch}{1}
\begin{array}{c}
\text{$\beta$-label}\\(0,\,4)
\end{array}}
&
\renewcommand{\arraystretch}{1}
\begin{array}{c}
\text{$\beta$-label}\\(2,\,2)
\end{array}
&
\renewcommand{\arraystretch}{1.5}
\phantom{
\begin{array}{c}
a\\b
\end{array}}
\\\hline
\renewcommand{\arraystretch}{1}
\begin{array}{c}
\text{$\gamma$-label}\\(0,\,4)
\end{array}
\phantom{\ }
&
\renewcommand{\arraystretch}{1.5}
\begin{array}{c}
\phantom{\ }E_4-E_1\phantom{\ } \\\Diamond > 0
\end{array}
&
\renewcommand{\arraystretch}{1.5}
\begin{array}{c}
\phantom{\ }E_4+E_1 \phantom{\ }\\ \Diamond < 0
\end{array}
&\phantom{\ }  \mathcal{O}_{\,\cdot\, K}
\\\hline
\renewcommand{\arraystretch}{1}
\begin{array}{c}
\text{$\gamma$-label}\\(2,\,2)
\end{array}
\qquad &
\renewcommand{\arraystretch}{1.5}
\begin{array}{c}
F_4+E_1 \\\Diamond \geq 0
\end{array}
\quad
&
\renewcommand{\arraystretch}{1.5}
\begin{array}{c}
F_4-E_1\\
\Diamond < 0
\end{array}&
\phantom{\ }\mathcal{O}_{\,\cdot\, K}'
\\\hline
&
\renewcommand{\arraystretch}{1.5}
\begin{array}{c}
\mathcal{O}_3
\end{array}
&
\begin{array}{c}
\mathcal{O}_4
\end{array}
&
\end{array}
\end{equation}
\caption{\small The four $\tilde{K}$-orbits $\mathcal{O}_{3K},\,\mathcal{O}_{3K}',\,\mathcal{O}_{4K},\,\mathcal{O}_{4K}'$ within the two $\rm{G}_{2(2)}$-orbits $\mathcal{O}_{3}$ and $\mathcal{O}_{4}$.
Our labeling of the orbits is indicated by the last row and the rightmost column.}
\label{korbitar}
\end{center}
\end{table}

We  now discuss a more physical way of distinguishing the orbits. It is well known that in $ N=8$ supergravity there exists an $\mathrm{E}_{7(7)}$
invariant quartic polynomial $\Diamond$ of the charges that is proportional to
the square of the entropy \cite{Kallosh:1996uy} of the corresponding
black hole. The S$^3$ model is a consistent truncation of $N=8$
supergravity. Therefore, the $\mathrm{E}_{7(7)}$ quartic polynomial descends to
a function of electromagnetic charges in the S$^3$ model. The quartic
invariant of this theory is invariant under the global
$\mathrm{SL}(2,\mathbb{R})$ symmetry of the theory.  In terms
of the $\rom{M2}^3$, $\rom{M5}^3$, $\rom{P}$, and $\rom{KKM}$ charges defined in section
\ref{sec:S3theory}, the polynomial reads
\begin{equation}
\label{eqn:quarticinvariant}
\Diamond(Q_1, Q_2, P_1, P_2) = 3(Q_2 P_2)^2+6 Q_1 P_1Q_2
P_2-(Q_1P_1)^2+4 Q_2^3 P_1+4 Q_1 P_2^3.
\end{equation}
It turns out that $\Diamond$ vanishes in the case of $\mathcal{O}_1$ and $\mathcal{O}_2$ and in general is non-zero for the remaining orbits. The particular relations are shown in Table \ref{korbitar}.

\subsection{Generating orbits in practice}

To generate the full orbits, we start with a nilpotent representative of an orbit and act on it with
$\tilde{K}$ by conjugation. Each element of the orbit generated in this way corresponds to an extremal solution. The coordinates on $\tilde{K}$ hence parameterize spacetime solutions. For this reason it is instructive to discuss the manifold $\tilde{K}$ in some detail. Recall that using the Iwasawa-decomposition
we can write $\tilde{K}$ as
\begin{equation}
\label{eqn:IwasawaofH}
\tilde{K}= \cK \cA \cN,
\end{equation}
where
$\cK = \mathrm{SO}(2) \times \mathrm{SO}(2)$
is the maximal compact subgroup of $\tilde K$, $\cA$ is an Abelian non-compact subgroup generated by the two Cartan elements in $\mathfrak{h}$, and
$\cN$ is a nilpotent subgroup generated by $E_2$ and $E_6$ in $\tilde{\mathfrak{k}}$ (or $F_2$ and $E_6$ as in the $\mathcal{O}_1$ case below).
The Iwasawa decomposition can also be written in the form
\begin{align} \label{iwasawajakob}
\tilde{K}= \cK \cA \cN=\cK (\cA \cN \cA^{-1})\cA = \cK\cN\cA.
\end{align}
It turns out that this form is more useful because we choose linear combinations of the root vectors as representatives of various nilpotent orbits.
Since root vectors are eigenvectors of the Cartan subalgebra $\mathfrak{h}$,
the action of $\cA$ only changes the coefficients in the linear combinations.
Thus, by keeping these coefficients unspecified we can simply omit the factor
$\cA$ in (\ref{iwasawajakob}) and only act with $\cK\cN$. This is how we generate the $\tilde K$-orbits below.
In terms of the electromagnetic and scalar charges defined earlier,  the most general charge matrix for the S$^3$ model is written as
\begin{align}
\mathcal{Q}&=-2M h_1+(-\Sigma-M)h_2-Q_1 p_1+\Xi p_2-Q_2 p_3+\tfrac12 P_2 p_4+\tfrac16 P_1p_5+\tfrac16 N p_6.\label{generalchargematrix}
\end{align}
In order to obtain
regular black holes, we require
the NUT charge to vanish ($N=0$) when generating the orbits below. This  implies that $\omega_3$ in \eqref{eqn:metric4dto3d} vanishes identically for all solutions  we consider.


\setcounter{equation}{0}
\section{Supersymmetric orbits}
\label{SupersymmetricOrbits}
In this section  we discuss supersymmetric orbits in more detail. For each orbit we present the most general solution and a few examples.

\subsection{The $\mathcal{O}_1$ orbit}
The  $\mathcal{O}_1$ orbit is the smallest of the nilpotent orbits.
It is characterized uniquely by charge matrices that square to zero,
\begin{equation}
\label{eqn:quadraticcharges}
{\cal Q}^2 = 0.
\end{equation}
As explained in section \ref{genG2} we can take
the representative for this orbit to be $E_1$, which is an element in $\tilde{\mathfrak{p}}$ and (unlike $E_5$) has vanishing $p_6$ coefficient. Thus $E_1$ corresponds to a charge matrix without  NUT charge. However, when we act on $E_1$ with
a general element in $\tilde K$, the property that the NUT charge vanishes is not preserved and it must be imposed by hand.

Since both $F_2$ and $E_6$ commute with $E_1$, the action of the nilpotent subgroup $\cN$ on $E_1$ is trivial. (This is the reason why we chose $F_2$ and $E_6$ rather than $E_2$ and $E_6$ as generators of $\cN$ in this case.)
Therefore, the full $\mathcal{O}_1$ orbit can simply be generated as
\be
\cQ = \mathrm{Ad} \left(a k_2 + b k_6\right) (4m E_1) \label{charge-matrix}
\ee
where $a,\,b$, and $m$ are arbitrary real parameters.
The general expression for the NUT charge in this orbit is
\be
N = - m \sin 12b,
\ee
and the no-NUT condition  becomes
\begin{align}
6 b &= \frac{j \pi}{2}, \quad j \in \mathbb{Z}.
\end{align}
Modulo redefinitions of $a$,
\be
a \rightarrow a + \frac{j\pi}{2} \label{shift},
\ee
the electromagnetic charges are given by
\begin{align}
Q_1 &= 2 m \sin ^3a , &  Q_2 &=  2 m \cos a \sin ^2a,\nn\\
P_1 &= -2 m\cos ^3a,  &  P_2 &= -2 m\cos ^2a \sin a.
\end{align}
For the mass and scalar charges,
it matters if  $j$ is even or odd. The two cases differ in the signs of these charges.
For  odd (even) $j$, we find
\begin{align}
M &= \pm m, & \Xi &= \pm   m \sin 2 a, & \Sigma &= \pm m \cos 2 a.
\end{align}
It is clear from these expressions that
the even and odd cases are related by $m \rightarrow -m$ and $a\rightarrow a + \pi$. Thus it suffices to consider only one case. We choose $j = 1$ and restrict the parameters to the range
\begin{align}
m &> 0,& 0\leq a &<2\pi. \label{range1}
\end{align}

To obtain the general form of the metric we exponentiate the charge matrix and read off the scalars from the resulting geodesic on the coset manifold. For example, the scalars $\phi_1, \phi_2$ and $\chi_1$ are given by
\begin{align}
e^{\frac{\phi_1}{\sqrt{3}}+\phi_2}  &= 1+ \frac{2 m \cos^2 a}{r}, & e^{2\phi_2} &= 1+ \frac{2m}{r}, &
\chi_1 &= \frac{2 m \sin^3a}{2 m+r}.
\end{align}
In the parameter range \eqref{range1} these exponentials are finite and positive definite.
Since we require  vanishing  NUT charge ($\omega_3 = 0$), the one-form $B_1$ takes the form
\begin{equation}
\label{eqn:B1}
B_1 = P_1 \cos \theta  d\phi.
\end{equation}
It is then straightforward to write the full metric from (\ref{eqn:canonicalform}).

We now show that the general solution of this orbit can be written manifestly in the Gibbons-Hawking form. To this end we perform the coordinate transformation \eqref{eqn:Zshift} with the choice
\begin{equation}
 \label{eqn:o1coord}
v =  \sin 3a.
\end{equation}
One finds
\begin{equation}
f = - \frac{r \cos 3a  + 2m \cos^3 a}{r + 2m \cos^2a},
\end{equation}
and\footnote{When $f < 0$ everywhere we interpret the solution by taking $ f \to -f, t \to -t, \omega \to -\omega$ and reversing the orientation of the base manifold \cite{Gauntlett:2002nw}. In certain examples discussed below we implicitly do this sign flipping to  present the solution.}
\begin{equation}
H = - \cos 3 a +  \frac{P_1}{r}.
\end{equation}
Note that for certain ranges of the parameter $a$ and the coordinate $r$ these functions become negative or zero. However, this is not a problem: even though the base space becomes `unphysical', the time fibration is such that the five-dimensional spacetime is well behaved. This situation is reminiscent of several well known smooth solutions (e.g., \cite{Giusto:2004kj, Bena:2005va, Berglund:2005vb}) where the base space is singular but the time fibration is such that the total spacetime is  smooth.

The one-form $\omega$ on the base space can be readily calculated,
\begin{align}
\omega &= \omega_5 (dz + P_1 \cos \theta d\phi), &\omega_5 &= \frac{1}{H^2}\left(- \sin 3 a + \frac{3P_2}{r}\right).
\end{align}
It follows from \eqref{eqn:B1} and \eqref{eqn:chi} that the harmonic function $H$ obeys the Gibbons-Hawking condition \eqref{eqn:GibbonsHawking}. It is straightforward to verify \eqref{eqn:chi2}--\eqref{eqn:B2}, confirming that the field strength is of the form
\eqref{eqn:FieldStrength}. The quartic invariant for this orbit is identically zero.

The $\mathcal{O}_1$ orbit contains
extremal Kaluza-Klein wave and Kaluza-Klein monopole among other solutions. We now list how they can be obtained from the general expressions for this orbit.
\subsubsection{Extremal Kaluza-Klein wave}
\label{sec:KKwave}
When $a= \pi/2$ and when $a=3\pi/2$,
the warp factor $f$ vanishes and we end up in the null-case, i.e., the case for which the Killing vector constructed from the Killing spinor is a null vector. These cases correspond to the positively and negatively charged extremal Kaluza-Klein waves, respectively.
To see this explicitly we set $m = {Q}/{2}, a = {\pi}/{2}$ to obtain the positively charged solution, and $m = {Q}/{2}, a = {3\pi}/{2}$ to obtain the negatively charged solution in the $\mathcal{O}_1$ orbit. The metric becomes
\begin{align}
ds^2_5 &= ds_{3}^2 - V^{-1} dt^2 + V\left(dz\pm\frac{Q}{V r}dt\right)^2, & V &=  1+\frac{Q}{r}, \qquad  Q> 0,
\end{align}
while the Maxwell field vanishes.
The non-zero scalars are
\begin{align}
\phi _1 &= -\frac{1}{2} \sqrt{3} \log V, &
\phi _2  &=
\frac{1}{2} \log V, &
\chi_1 &= \pm \frac{Q}{Q+r}.
\end{align}
and the non-zero charges are
\begin{align} \label{charges-kkp}
Q_1 &= \pm Q, &
M &= \frac{Q}{2},&
\Sigma &= - \frac{Q}{2}.
\end{align}

\subsubsection{Extremal Kaluza-Klein monopole}
\label{sec:KKmonopole}
The other special cases are when $a=0$ and $a = \pi$. Then the coordinate transformation \eqref{eqn:Zshift}, with \eqref{eqn:o1coord} inserted, does nothing. These are the  negatively and positively charged extremal Kaluza-Klein monopoles, respectively. By setting $m = {P}/{2}, a = 0$ we obtain the negatively charged solution, and by setting $m = {P}/{2}, a = \pi$ we obtain the positively charged solution. We find that
the metric is expressed as
\begin{align}
ds_{5}^2 &= -dt^2 + H ds_{3}^2+ H^{-1} (dz \mp P \cos \theta \, d\phi)^2, & H &= 1+\frac{P}{r}, & P &> 0,
\end{align}
while the Maxwell field vanishes.
It follows that the non-zero scalars are given as
\begin{align}
\phi_1 &= \frac{1}{2} \sqrt{3} \log H, & \phi_2 &= \frac{1}{2} \log H,& \chi_5 &= \mp \frac{P}{P+r},
\end{align}
and the non-zero charges as
\begin{align} \label{charges-kkm}
P_1 & = \mp P, &
M &= \frac{P}{2}, &
\Sigma &= \frac{P}{2}.
\end{align}

\subsection{The $\mathcal{O}_2$ orbit}
The second smallest orbit is the $\mathcal{O}_2$ orbit. We generate this orbit by acting with $\tilde{K}$ on $E_4$. Since $E_6$ commutes with $E_4$, the action of the nilpotent subgroup generated by $E_6$ is trivial.
Therefore, the full $\mathcal{O}_2$ orbit can simply be generated by
\be
\cQ = \left(\mathrm{Ad} \left(a k_2 + b k_6\right) \circ \mathrm{Ad}(c E_2)\right)(m E_4), \label{ChargeMatrixO2}
\ee
where $a,\,b,\,c$, and $m$ are arbitrary real parameters.
The general expression for the NUT charge in this orbit is
\be
N = \frac{3m}{2} (c \cos 12 b+\sin 12 b),
\ee
yielding that the no-NUT condition becomes
\begin{align}
\label{eqn:o2nonut}
&\cos 12 b \neq 0, \qquad c = - \tan 12b .
\end{align}
When $ \cos 12 b= 0 $ the no-NUT condition has no solution. With $c$ given by (\ref{eqn:o2nonut})
we find
\begin{align}
M &=  \tfrac{3}{2}n,\nn\\
Q_1 &= 3n  \cos^2 \beta\sin (\alpha + \beta),\nn\\
P_1 &=-3n  \sin^2 \beta\cos (\alpha + \beta),\nn\\
Q_2 &=\tfrac{1}{4}n \left[3\cos(-3\beta-\alpha)+2\cos(\alpha+\beta)-\cos (\beta-\alpha)\right],\nn\\
P_2 &= -\tfrac{1}{4}n \left[3\sin(-3\beta-\alpha)+2\sin(\alpha+\beta)-\sin (\beta-\alpha)\right],\label{generalO2}
\end{align}
and
\begin{align}\label{generalO2-1}
\Sigma &= \tfrac{1}{2}n  \left(\cos 2 (\alpha + \beta)  -2 \cos 2 \beta\right),\nn\\
\Xi &= \tfrac{1}{2}n \left(  \sin 2(\alpha + \beta) - 2 \sin 2 \beta \right),
\end{align}
where $\alpha = 12 b$, $\beta = a - 6 b$ and
$n={m}/{\cos \alpha}$.
The expressions (\ref{generalO2}) and \eqref{generalO2-1} are invariant under
\begin{align}
m &\rightarrow -m, & \alpha &\rightarrow \alpha + \pi ,& \beta &\rightarrow \beta + \pi.
\end{align}
We now restrict the parameters to the range
\begin{align}
m &> 0,& - \frac{\pi}{2}&<\alpha<\frac{\pi}{2},&
0 &\leq \beta<2\pi,
\label{rangej=1}
\end{align}
to ensure that the mass is finite and positive definite. To obtain the general form of the metric, we exponentiate the charge matrix and read off the scalars from the resulting geodesic on the coset manifold. We find
\be
e^{\frac{\phi_1}{\sqrt{3}}+\phi_2}  = 1+\frac{ \Sigma + M  }{ r}+ \frac{m^2 \sin ^2\beta}{r^2},
\ee
and
\be
e^{2\phi_2} = 1+\frac{2M}{r}+ \frac{3 m^2}{r^2}+  \frac{m^3 \cos \alpha}{r^3}.
\ee
Note that in the parameter range \eqref{rangej=1} these exponentials are finite and positive definite.  The scalar field $\chi_1$ takes the form
\begin{align}
\chi_1 &=  e^{-2\phi_2}\Bigg(  \frac{Q_1}{r}
+  \frac{3  m^2 \cos  \beta  \sin (\alpha +2 \beta ) }{r^2} \nn \\ & \qquad \qquad \qquad  +  \frac{ m^3 (3 \sin  \beta +\sin  3 \beta
 +2 \sin  (2 \alpha +3 \beta ))}{4 r^3} \Bigg).
\end{align}
The full metric (\ref{eqn:canonicalform}) is then easily obtained by requiring
the no-NUT charge condition ($\omega_3 = 0$) which leads to the one-form $B_1$, \begin{equation}
\label{eqn:B1o2}
B_1 = P_1 \cos \theta  d\phi.
\end{equation}

This orbit contains in particular the $\mathrm{M2}^3$ and the $\mathrm{M5}^3$ solutions, which we will discuss below. Now we show that the general solution in this orbit can also be written in the Gibbons-Hawking form. With the coordinate transformation \eqref{eqn:Zshift}
\begin{equation}
 \label{eqn:o2coord}
v =  - \sin ( \alpha + 3 \beta),
\end{equation}
one finds
\begin{equation}
f =  H e^{-\frac{\phi_1}{\sqrt{3}}-\phi_2},
\end{equation}
and
\begin{equation}
H =  \cos ( \alpha + 3 \beta) +  \frac{P_1}{r},
\end{equation}
together with
\be
\omega_5 = \frac{1}{H^2}\left(
\sin (\alpha +3 \beta )+\frac{3 P_2}{r}+  \frac{3 m^2 \cos (\alpha +2 \beta ) \sin \beta}{r^2}-\frac{m^3 \sin ^3\beta }{r^3}\right).
\ee
The Maxwell field is again of the form \eqref{eqn:FieldStrength}. Observe that
the null case is obtained when $H=0$, i.e., when $\cos(\alpha+3\beta) = 0$ and $P_1 = 0$. This implies $(i)$ $\alpha = 0$, $\beta = {3\pi}/{2}$ or $(ii)$ $ \alpha = 0$, $\beta = {\pi}/{2}$. These cases precisely correspond to the positively and negatively charged $\mathrm{M5}^3$ solutions respectively. As in the $\mathcal{O}_1$ case, the quartic invariant for this orbit is identically zero.

\subsubsection{Extremal $\rom{M2}^3$}
The positively (negatively) charged  extremal $\rom{M2}^3$ solution is obtained by taking $m = Q$, $\alpha = 0$, $ \beta = 0$ ($m = Q, \alpha = 0$, $\beta = \pi$) in the $\mathcal{O}_2$ orbit. These solutions are
\begin{align}
ds^2_5 &= -f^2 dt^2 + f^{-1}(dz^2 + ds_3^2), & f^{-1} &= 1+\frac{Q}{r}, & Q &> 0,\nn\\
A&= \pm \sqrt{3}\left(1-f\right) dt. &&
\end{align}
The non-zero scalars are
\begin{align}
   \phi_1 &= \frac{\sqrt{3}}{2}  \log  f, &\phi_2  &=   -\frac{3}{2} \log f,&\chi_3 &=  \pm \sqrt{3}\left(1-f\right).
\end{align}
and the non-zero charges are
\begin{align} \label{charges-m23}
 Q_2 &= \pm Q, &  M &= \frac{3 Q}{2},& \Sigma &= -\frac{Q}{2}.
\end{align}

\subsubsection{Extremal $\rom{M5^3}$}

The positively (negatively) charged  extremal $\rom{M5}^3$ solution is obtained by taking $m = P$, $\alpha = 0$, $ \beta ={3\pi}/{2}$ ($m = P, \alpha = 0$, $\beta = {\pi}/{2}$) in the $\mathcal{O}_2$ orbit. These solutions are
\begin{align}
ds^2_{5} &= V^{-1} \left( - dt^2 +  dz^2 \right) + V^2 ds^2_3, & V &= 1 + \frac{P}{r},&P&>0,\nn\\
A &= \mp \sqrt{3} P \cos \theta d\phi.&&
\end{align}
The non-zero scalars are
\begin{align}
\phi_1 &= \frac{\sqrt3}{2} \log V,&
\phi_2  &= \frac{3}{2} \log V,& \chi_4 &= \mp \sqrt{3} \left(1 - V^{-1}\right),
\end{align}
and the non-zero charges are
\begin{align} \label{charges-m53}
P_2 &= \pm P,  & M&= \frac{3 P}{2}, & \Sigma &= \frac{P}{2}.
\end{align}

\subsection{The $\mathcal{O}_{3K}$ orbit}
\label{sec:O3}
The first of the bigger orbits is a four parameter family of supersymmetric black holes. This orbit corresponds to the BPS attractor of \cite{Gaiotto:2007ag}. It is characterized by positive values of the quartic invariant $\Diamond$. One may generate the full orbit  by acting with $\tilde{K}$ on $E_4 - E_1$. However, it is more convenient to be a little more general and start with the arbitrary combination $n E_1+m E_4$ ($n < 0,\,m > 0$). The full $\mathcal{O}_{3K}$ orbit is then generated by
\be
\cQ = \left(\mathrm{Ad} \left(a k_2 + b k_6\right) \circ \mathrm{Ad}(c F_2)\right)\left(nE_1 + mE_4\right). \label{ChargeMatrixO3}
\ee
The
NUT charge in this orbit is given by
\be
N = \frac{1}{4}\left[-12 c\,m\, \cos 12 b-\left(n+6 \left(c^2-1\right) m\right) \sin 12 b\right],
\ee
and thus
the no-NUT condition is
\begin{align}
n&= -6 m \left(c^2+2 c \cot 12 b -1\right),& \mbox{when}~& \sin 12 b \neq 0,  \label{eqn:nonutO3} \\
c&=0, &  \mbox{when}~&\sin 12 b = 0.\label{eqn:nonutO3-1}
\end{align}
Taking into account the no-NUT condition and using the parameterization $\alpha = 12 b $ and $\beta = a - 6 b$   we get the mass and the electromagnetic charges to be
\begin{align}
M&= 3 md,\nn\\
Q_1 &= -3m \sin \beta \left[2 d \sin \beta \sin(\beta + \alpha)
-1\right],\nn\\
P_1 &= 3m \cos \beta \left[2 d \cos \beta \cos(\beta + \alpha)
-1\right],\nn\\
Q_2 &= -\tfrac12 m \left[d \cos (\beta-\alpha)-2 \cos \beta +2 d \cos (\alpha +\beta)
-3 d \cos (-\alpha -3 \beta )\right],\nn\\
\label{generalO31}
P_2 &= \tfrac12 m \left[d \sin (\beta -\alpha)-2 \sin \beta+2 d \sin (\alpha +\beta )
-3 d \sin (-\alpha -3 \beta )\right],
\end{align}
and the scalar charges to be
\begin{align}
\Sigma &= m  \left[2 d \cos2 \beta+d \cos (2 (\alpha +\beta ))-2 \cos (\alpha +2 \beta )\right],\nn\\
   \Xi &= m \left[2 d \sin2 \beta+d \sin (2 (\alpha +\beta ))-2 \sin (\alpha +2 \beta )\right],
\label{generalO32}
\end{align}
where $d$ is chosen such that
 \begin{align}
 d&= \frac{c}{\sin \alpha} \qquad \qquad \qquad \quad  \, \mbox{when~}\sin\alpha \neq 0 \label{eqn:dsub},\\
d &= \frac{1}{2}\left(1-\frac{n}{6m}\right) \qquad \qquad \mbox{when~}\sin\alpha = 0  .
\end{align}
In terms of these parameters $m, \alpha$, and $d$, the quartic invariant  in this orbit is given by
\begin{equation}
\Diamond =  12\,m^4 \left(d^2 \sin^2 \alpha +2 d \cos \alpha -1\right).
\end{equation}
It follows
from (\ref{eqn:nonutO3}) and \eqref{eqn:nonutO3-1} that $\Diamond>0$ as long as $n < 0$ and $m>0$.
The expressions (\ref{generalO31}) and (\ref{generalO32}) are invariant under
\begin{align}
m &\rightarrow -m, & d&\rightarrow -d,  & \alpha &\rightarrow \alpha + \pi , &\beta &\rightarrow \beta + \pi,
\end{align}
We now restrict the parameters to the range
\begin{align}
m &> 0, &\quad d&> 0,&
0 &\leq \alpha, \beta<2\pi.  \end{align}
to ensure that the mass is  positive definite.
As explained for the $\mathcal{O}_{1}$ and $\mathcal{O}_{2}$ orbits, obtaining the general form for the metric is straightforward: it can be read off from the resulting geodesic on the coset manifold by exponentiating the charge matrix.
The spacetime solution is parameterized in an unilluminating form. To give an idea to the reader how these expressions look like we write out the
dilatons:
\begin{align}
 e^{\frac{\phi_1}{\sqrt{3}} +  \phi_2} &= 1 + \frac{\Sigma + M}{r} + \frac{k}{r^2},\\
e^{2\phi_2} &= \frac{1}{r^4}\left(\frac{\Diamond}{4}  + c_1 r + c_2 r^2 + 2 M r^3 + r^4
 \right),\label{o3warp}
\end{align}
where
\begin{align}
k = \,&m^2 \left(6 d \cos \alpha \cos ^2\beta +2 d^2 \sin ^2\alpha +2 \cos 2 \beta  (d^2 \sin ^2\alpha -1)+d \sin \alpha \sin 2 \beta-1\right),\nn\\
c_1 = \,&2 m^3 \left(4 d^3 \sin ^4\alpha +12 d (d \cos \alpha -1) \sin ^2\alpha +9 d-4 \cos \alpha \right), \nn\\
c_2 =\,&6 m^2 \left(2 d^2\sin^2 \alpha +3 d\cos \alpha -1\right).
\end{align}
The one-form $B_1$ takes  the form
\begin{equation}
\label{eqn:B1o3}
B_1 = P_1 \cos \theta  d\phi.
\end{equation}
Since we have set the NUT charge to zero we have $\omega_3 = 0$ in (\ref{eqn:canonicalform}).

This orbit contains in particular the $\mathrm{M2}^3-$KKM and the $\mathrm{M2}^3-\mathrm{M5}^3-$P solutions, which we will discuss below. We now show that the general solution in this orbit can be written in the Gibbons-Hawking form. To this end we perform the coordinate transformation \eqref{eqn:Zshift} with the choice
\begin{equation}
 \label{eqn:o3coord}
v =  - \sin ( \alpha + 3 \beta).
\end{equation}
One finds
\begin{equation}
f =  H e^{-\frac{\phi_1}{\sqrt{3}}-\phi_2},
\end{equation}
with
\begin{equation}
H =  \cos ( \alpha + 3 \beta) +  \frac{P_1}{r}.
\end{equation}
Finally, we have
\begin{equation}
\omega_5 = \frac{1}{H^2}\left(\sin(\alpha+3 \beta) + \frac{3 P_2}{r}+ \frac{b_2}{r^2}+ \frac{b_3}{r^3}\right),
\end{equation}
where $b_2$ and $b_3$ are complicated trigonometric functions which are not particularly enlightening and so we omit them.
The Maxwell field is again of the form \eqref{eqn:FieldStrength}.

Since the above analysis is also valid for $n> 0$, we have now in fact
shown that the full $\mathcal{O}_{4K}$ orbit can also be written in
the Gibbons-Hawking form. However, it is
easy to see that the $\mathcal{O}_{4K}$ orbit is unphysical---a generic solution in this
orbit contains singularities outside of the horizon. To see this we
look at the four-dimensional warp factor (\ref{o3warp}).
For the
$\mathcal{O}_{4K}$ orbit $\Diamond < 0$, therefore the warp factor approaches negative infinity as
$r$ goes to zero.  For both the $\mathcal{O}_{3K}$ and $\mathcal{O}_{4K}$ orbits the warp factor (\ref{o3warp}) is unity at spatial infinity.  This implies that the warp factor becomes
zero for the $\mathcal{O}_{4K}$ orbit at least at one point outside of  the horizon. At this point the
spacetime is singular. {The singularity is not a coordinate singularity but a genuine curvature singularity. This can be most easily seen by looking at  curvature invariants, e.g.,  $R_{\mu \nu \rho \sigma} R^{\mu \nu \rho \sigma}$, for the four-dimensional metric (\ref{eqn:metric4dto3d})--(\ref{3dmetric}) with $\omega_3 =0$. We see that $R_{\mu \nu \rho \sigma} R^{\mu \nu \rho \sigma}$ diverges as $\phi_2$ tends to minus infinity.
Thus, we conclude that the $\mathcal{O}_{4K}$ orbit is unphysical.}

\subsubsection{Two parameter $\rom{M2^3-KKM}$ family}\label{ssec:M2KKM}
Now we discuss examples in the $\mathcal{O}_{3K}$ orbit. The first example is a supersymmetric static electrically charged black hole sitting at the center of the Taub-NUT. The solution takes the form
\begin{align}
ds^2_5 &= -f^2 dt^2 + f^{-1}H^{-1} (dz +P \cos \theta d\phi)^2 + f^{-1} H ds^2_3,
\\
A &= \sqrt{3} \left( 1-f\right) dt,
\end{align}
where
\begin{align}
f^{-1} &= 1+ \frac{Q}{r}, &  H &= 1 + \frac{P}{r}, & P&> 0, & Q &> 0.
\end{align}
We use $Q$ to denote the $\rom{M2^3}$ charge and $P$ to denote the KKM charge. There are several ways to obtain this solution. It can be obtained by setting $J=0$ in the solution discussed in the appendix of \cite{Gaiotto:2005gf}, or by setting $R=0,~q=0$ in the black ring solution discussed in \cite{Elvang:2005sa}, or by setting three charges equal in the familiar four charge black hole in four dimensions. An important property of this solution is that upon setting $P=Q$ it reduces to an extremal four-dimensional Reissner-Nordstr\"om solution. The non-zero scalars for this solution are
\begin{align}
\phi_1 &=\frac{\sqrt{3}}{2}  \log \left( fH\right), &\phi_2 &=\frac{1}{2}\log\left(H f^{-3}\right),\nn \\
\chi_3 &= \sqrt{3}\left(1-f\right), &\chi_5 &= 1-H^{-1}.
\end{align}
The non-zero charges are
\begin{align} \label{M2kkmcharg1}
\Sigma &= \frac{P- Q }{2}, & M &= \frac{P+3  Q }{2}, \nn \\
Q_2 &= Q, & P_1&= P.
\end{align}
The quartic invariant for the solution is $\Diamond = 4P Q^3$. In the above parameterization of the $\mathcal{O}_{3K}$ orbit, this solution can be obtained by setting
\begin{align}
 m &=  Q, & d &= \frac{1}{2}\left( 1 + \frac{P}{3Q}\right), &\alpha& = \beta = 0.
\end{align}

\subsubsection{Two parameter $\rom{M2^3-M5^3-P}$ family}
\label{sec:Bena}
The second example we discuss
is the black string of
\cite{Bena:2004wv}. It is a bound state of $\rom{M5}^3$ with $\rom{M2}^3$ and a critical amount of KK momentum along the string so that it corresponds to the infinite radius limit of the supersymmetric black ring \cite{Elvang:2004rt} of minimal supergravity. The solution is
\begin{align}
ds^2_5 &= -f^2 (dt + \omega_z dz)^2 + f^{-1}(dz^2 + ds^2_3),\\
A&=-\frac{\sqrt{3}}{2} \, q \, \left(1 +\cos\theta \right)d\phi +
\sqrt{3}\left(1-f\right) dt - \sqrt{3} \left( \omega_z f
-\frac{q}{2r}\right)dz,
\end{align}
where
\begin{align}
\omega_z &= \frac{3 q}{2 r}+\frac{3 Q q}{4
r^2}+ \frac{q^3}{8 r^3}, & f^{-1} &= 1+\frac{Q}{r}+\frac{q^2}{4 r^2}, & q & > 0, & Q > 0.\end{align}
The non-zero charges are
\begin{align}
\label{charges-m23m53p}
Q_2 &= Q,& P_2 &= \frac{q}{2}, &  Q_1 &=
-\frac{3 q}{2},\nn\\
M &= \frac{3 Q}{2},& \Sigma &= -\frac{Q}{2}, &\Xi &= -q,
\end{align}
and the quartic invariant is given by
\begin{equation}
\Diamond = \frac{3}{4} q^2(Q^2-q^2).
\end{equation}
Thus, the solution is smooth and supersymmetric for $Q > q$. It can be obtained by setting
\begin{align}
m  &= \frac{q}{2},& d &= \frac{Q}{q}, &  \alpha &= \beta = \frac{3 \pi}{2},
\end{align}
in the general expressions of the $\mathcal{O}_{3K}$
orbit given above.

\subsubsection{Three parameter $\rom{M2^3-M5^3-P}$ family}
\label{sec:susystring}
A three parameter family of supersymmetric black strings of minimal supergravity with three independent $\rom{M2^3-M5^3-P}$ charges is also contained in the $\mathcal{O}_{3K}$ orbit. {In the four-dimensional language this solution is previously known in the literature \cite{Shmakova}. It} is most conveniently parameterized by setting
\begin{align}
m &= \frac{q}{2}, & d &= \frac{\sqrt{Q^2+\Delta ^2}}{q}, &\alpha &= -\arccos \frac{\Delta }{\sqrt{Q^2+\Delta ^2}}, & \beta &= \frac{3\pi}{2},
\end{align}
with $q>0$ and $Q>0$ in the general expressions for this orbit.
The metric and the gauge field (after an appropriate rescaling of the $t$- and $z$-coordinates in the Gibbons-Hawking form of the  $\mathcal{O}_{3K}$ orbit) are:
\begin{align}
\label{string3}
ds^2_5 &= - f^2 (dt + \omega_z dz)^2 + f^{-1} (dz^2 + ds_3^2),\\
\label{maxwell3}
A &= -\frac{ \sqrt{3}}{2} q(1 +  \cos \theta) d\phi+ \sqrt{3} \left(1-f\right)dt \nn \\
 &\quad\, - \sqrt{3} \left(f \left(\omega_z - \frac{\Delta}{S} - \frac{\Delta (Q^2 + q\Delta)}{S^2 r } - \frac{q^2\Delta}{4 S r^2}\right)- \frac{q}{2r}  \right)dz,
\end{align}
with
\begin{align}
f^{-1} &= 1+\frac{Q^2+q \Delta}{S r}+ \frac{q^2}{4 r^2},\\
\omega_z &=
\frac{\Delta }{S} + \frac{3 q}{2 r}+\frac{3 \left(\Delta  q^2+Q^2 q\right)}{4 r^2 S} + \frac{q^3}{8 r^3},
\end{align}
and $S = \sqrt{Q^2 + \Delta^2}$.
The non-zero charges for this family are
\begin{align}
Q_2 &= Q, & P_2 &= \frac{q}{2}, & Q_1 &= 3 \Delta -\frac{3 q}{2},\nn\\  M
&= \frac{3}{2} S,&
\Sigma &= -\frac{S^2 + 2 \Delta (\Delta -q)}{2 S}, & \Xi &= \frac{Q (\Delta -q)}{S}.
\end{align}
The quartic invariant is
\be
\Diamond = \frac{3}{4} q^2 \left(Q^2 -q^2+2 \Delta  q\right).
\ee
It is positive for $2\Delta \ge q$ or for $Q > \sqrt{q^2 - 2 q \Delta}$ when $2\Delta < q$. This expression for the quartic invariant can also be obtained \cite{Larsen:2005qr}
from the Maldacena-Strominger-Witten CFT \cite{Maldacena:1997de}.
 A construction of the non-extremal solution describing thermal excitations above this three parameter family was  outlined
in \cite{Compere:2009zh}. Roughly speaking, the three parameter family can be regarded as the boosted generalization of the two parameter family of section \ref{sec:Bena}.
The pressure density of the string \eqref{string3}---the $T_{zz}$ component of the ADM stress tensor (see e.g. section 2.1 of \cite{Myers:1999psa})---is
\be
T_{zz} = M + 3 \Sigma = \frac{3 (q-\Delta ) \Delta }{\sqrt{Q^2+\Delta ^2}}.
\ee
When $\Delta =0$ or $\Delta = q$ the string becomes pressureless.
Setting $\Delta = 0 $ the solution  \eqref{string3}--\eqref{maxwell3} reduces to the pressureless two parameter $\rom{M2^3-M5^3-P}$ family
of section \ref{sec:Bena}. Setting $\Delta = q$ one obtains
another two parameter family of supersymmetric pressureless black strings. In this family one can set $Q=0$ while keeping $q\neq 0$. This cannot be done in
the two parameter $\Delta =0$ family. Upon setting $Q=0$ in the $\Delta = q$ family one recovers the supersymmetric black string that describes the infinite radius limit of the
extremal singly spinning dipole ring of \cite{Emparan:2004wy}.
To the best of our knowledge the two parameter $\Delta = q$ family has not been previously
discussed in the literature.\footnote{Incorporating the no-NUT condition in the extremal limit of the black string discussed in appendix D of \cite{Elvang:2004xi}, one obtains a one parameter family in minimal supergravity. It corresponds to
$Q=0$ in our $\Delta = q$ family.} Perhaps it describes a novel supersymmetric limit of the infinite
radius limit of the conjectured five parameter non-extremal black ring \cite{Elvang:2004xi} in this theory. This
point certainly deserves further investigation. It is expected that the $\Delta = 0$ and $\Delta = q$ families are connected by a pressureless non-extremal black string \cite{Compere:2010fm}.

\subsubsection{Other examples}
One may obtain numerous other examples from the general expressions for the $\mathcal{O}_{3K}$ orbit presented above. In cases of sufficient complexity it is difficult to find parameter redefinitions that make the meaning of the parameters transparent from the spacetime point of view. However, it is relatively straightforward to check if a given  extremal solution belongs to the $\mathcal{O}_{3K}$ orbit. This simply amounts to confirming that the charge matrix for the given solution is nilpotent and the quartic invariant is positive definite. We checked the nilpotency of the charge matrix and the positivity of the quartic invariant for the $J \neq 0$ solution of \cite{Gaiotto:2005gf}\footnote{This calculation was also performed in \cite{Berkooz:2008rj}.} as well as for the $R=0$ solution of \cite{Elvang:2005sa}. Hence both of these configurations belong to the $\mathcal{O}_{3K}$ orbit. It would be interesting to find appropriate parameter redefinitions (and restrictions) in the $\mathcal{O}_{3K}$ orbit that precisely match with the parameterizations of the $J \neq 0$ solution as given in \cite{Gaiotto:2005gf} and the $R=0$ solution as given in \cite{Elvang:2005sa}.

\subsubsection{Truncation to the Einstein-Maxwell theory}
\label{sec:EinsteinMaxwell}
As mentioned  in section \ref{sec:S3theory}, the S$^3$ model admits a consistent truncation to minimal $N=2$ supergravity, i.e., to the Einstein-Maxwell theory. When reduced to three dimensions over a timelike Killing vector, the Einstein-Maxwell theory gives rise to the coset $\mathrm{SU}(2,1)/( \mathrm{SL}(2,\mathbb{R}) \times \mathrm{U}(1))$.
From the point of view of the three-dimensional hidden symmetry groups,
the consistent truncation is equivalent to choosing
the appropriate $\mathrm{SU}(2,1)$ subgroup of $\mathrm{G}_{2(2)}$. This can be done as follows.  The coset structure of minimal $N=2$ supergravity gives the reductive decomposition
\begin{equation}
\mathfrak{su}(2,1) = \mathfrak{sl}(2, \mathbb{R}) \oplus \mathbb{R} u \oplus \tilde{\mf{p}}',
\end{equation}
where $u$ generates a (compact) $\mathrm{U}(1)$ subgroup and $\tilde{\mf{p}}'$ is the $\bf 2_+ +2_-$ representation of  $\mathfrak{sl}(2, \mathbb{R}) \oplus \mathbb{R} u $.
If we denote the generators of  $\mathfrak{sl}(2, \mathbb{R})$ by $e', f',$ and $h'$, then
the compact subgroup generated by $e'-f'-u$ rotates the mass and NUT charge \cite{Houart:2009ed}.  The subgroup generated by $e'-f'+u$ rotates the electric and magnetic Maxwell charges \cite{Houart:2009ed}. Finding generators
for $\tilde{{\mf{k}}}$ that act in this way on $\tilde{\mf{p}}$ uniquely fixes an $\mathrm{SU}(2,1)$ subgroup inside $\mathrm{G}_{2(2)}$. This embedding also relates electromagnetic charges of the Einstein-Maxwell theory to those of the S$^3$ model.

More explicitly, we take the $\mathfrak{sl}(2, \mathbb{R})$ subalgebra to be spanned by $E_6,\,F_6$ and $H_6$, and let $K_2=E_2-F_2$ span the $\mathfrak{u}(1)$ subalgebra which commutes with  the $\mathfrak{sl}(2, \mathbb{R})$ subalgebra. It follows that the space $\tilde{\mf{p}}'$
is spanned by
\begin{align}
&E_1-\tfrac12 E_4, &&E_3-\tfrac16 E_5, &&  F_1-\tfrac12 F_4,  && F_3-\tfrac16 F_5.
\end{align}
Looking at the most general charge matrix spanned by these generators and comparing it with (\ref{generalchargematrix}) we immediately see that the generators corresponding to the scalar charges do not belong to $\mathfrak{su}(2,1)$. This is simply explained by the fact that
the Einstein-Maxwell theory contains no scalar fields. Furthermore, we find that the embedding of Einstein-Maxwell theory in the S$^3$ model is given by
\begin{align}
Q_2 &= P_1, &Q_1 &= P_2, & \Sigma  &= 0 ,& \Xi& =  0.
\label{eqn:EMembedding}
\end{align}
This is indeed
consistent with \eqref{EMtruncation}. It follows that
\be
\Diamond_\rom{EM}
 = (P_1^2 + Q_1^2)^2.
\ee
Hence, the Reissner-Nordstr\"om solution of the Einstein-Maxwell theory is naturally embedded into the $\mathcal{O}_{3K}$ orbit. It was shown in \cite{Houart:2009ed} that the Einstein-Maxwell theory has only one orbit of BPS solutions.


\setcounter{equation}{0}
\section{Non-supersymmetric orbit}
\label{NonSupersymmetricOrbits}
Having discussed the supersymmetric orbits, we now turn to the remaining two orbits, $\mathcal{O}_{3K}'$ and $\mathcal{O}_{4K}'$.  We will argue below that like $\mathcal{O}_{4K}$ the $\mathcal{O}_{3K}'$ orbit is also unphysical. The fact that for  $\mathcal{O}_{3K}$ and $\mathcal{O}_{4K}'$ the $\beta$-labels and $\gamma$-labels are the same is consistent with the observation of \cite{Bossard:2009we} that for physical orbits these labels should coincide. In the rest of the section we present general expressions for all charges and discuss examples in the $\mathcal{O}_{4K}'$ orbit.

\subsection{The $\mathcal{O}_{4K}'$ orbit}
\label{eqn:O4Kprime}

The $\mathcal{O}_{4K}'$ orbit is a four parameter family of non-supersymmetric black holes. This orbit
corresponds to the non-BPS attractor of \cite{Gaiotto:2007ag}. It is
characterized by negative values of the quartic invariant $\Diamond$. One may
generate the full orbit  by acting with $\tilde{K}$ on
$n E_1+ m F_4$ ($n<0,\,m>0$),
\be
\cQ = \left(\mathrm{Ad} \left(a k_2 + b k_6\right) \circ \mathrm{Ad}(c
F_2)\right)\left[n E_1+ m F_4\right]. \label{ChargeMatrixO4}
\ee
The general expression for the NUT charge of this orbit is
\be
N = \frac{1}{4} \Big((6\, m-n) \sin 12 b\,-\,6\, m \,c \cos 12 b \Big)
\ee
and
the no-NUT condition becomes
\begin{align}
n &= 6 m (1 - c \cot12b) & \mbox{when~} \sin 12 b \neq 0 \label{eqn:nonutO4},\\
c&=0 &\mbox{when~} \sin 12 b = 0.
\end{align}
Taking into account the no-NUT condition
and
using the parameterization $\alpha = 12 b $ and $\beta = a - 6 b$,   we
get the mass and electromagnetic charges to be
\begin{align}
M &= \tfrac{3}{2}  m d, \nn\\
Q_1 &= -3m \sin(\alpha + \beta) \left[\cos(\alpha + \beta) \cos \beta  - \sin(\alpha + \beta) \sin \beta +   d \sin^2(\alpha + \beta)\right],\nn\\
P_1 &= 3m \cos(\alpha + \beta) \left[\sin(\alpha + \beta) \sin \beta  - \cos(\alpha + \beta) \cos \beta +
     d \cos^2 (\alpha + \beta)\right],\nn\\
Q_2 &= \tfrac{1}{2} m \left[\cos \beta -6 d \cos (\alpha +\beta )
\sin ^2(\alpha +\beta )-3\cos (-2 \alpha -3 \beta )\right],\nn\\
P_2 &= - \tfrac{1}{2} m \left[\sin \beta -6 d \sin (\alpha +\beta )
\cos ^2(\alpha +\beta )-3\sin (-2 \alpha -3 \beta )\right],\label{paraO4M}
\end{align}
and the scalar charges to be
\begin{align}
\Sigma &= \tfrac{3}{2} m  d  \cos 2 (\alpha +\beta )-2 m  \cos (\alpha
+2 \beta ), \nn\\
\Xi & =   \tfrac{3}{2} m  d  \sin 2 (\alpha +\beta )-2 m  \sin (\alpha
+2 \beta ),\label{paraO4Xi}
\end{align}
where $d$ is given by
\begin{align}
d &= \frac{c}{\sin \alpha} \qquad \quad  \qquad \: \mbox{when~} \sin \alpha \neq 0,\label{dO4}\\
d &= \frac{1}{2}\left(1-\frac{n}{6m}\right) \qquad \mbox{when~} \sin \alpha = 0.\label{dO4}
\end{align}
The quartic invariant for these charges is
\begin{equation}
\Diamond =  -12 m^4 \cos ^2\alpha (d \cos \alpha-1),
\end{equation}
and is strictly negative provided $\cos \alpha \neq 0$ and $n < 0$. When $\cos \alpha = 0$, then $n=6m$ and we are no longer in the $\mathcal{O}_{4K}'$ orbit. Thus, for the  $\mathcal{O}_{4K}'$ orbit the quartic invariant is strictly negative.
The expressions (\ref{paraO4M}) and (\ref{paraO4Xi}) are invariant under
\begin{align}
m &\rightarrow -m, & d&\rightarrow -d,  &\alpha &\rightarrow \alpha + \pi ,  &\beta &\rightarrow \beta + \pi.
\end{align}
We now restrict the parameters to the range
\begin{align}
m &> 0, &d&> 0,&
0 &\leq \alpha,\beta<2\pi.
\end{align}
In this range the mass is positive definite. From (\ref{paraO4M}) and (\ref{paraO4Xi}) the general charge matrix  for the $\mathcal{O}_{4K}'$ orbit can be readily constructed.  To obtain the general form of the metric we exponentiate the charge matrix and read off the scalars from the resulting geodesic on the coset manifold. Again, the spacetime solution is parameterized in
an unilluminating manner. To give an idea to the reader how these
expressions look like we write out the four-dimensional warp factor:
\be
e^{2 \phi_2}= \frac{1}{r^4} \left( -
\frac{\Diamond}{4}  + 2 M r^3 +  c_1 r  + c_2 r^2 + r^4  \right)\label{warpO4},
\ee
where
\begin{align}
c_1 &=   m^3 \cos \alpha (9 d \cos \alpha -8), & c_2 &= 3 m^2 (3
d \cos \alpha-2).
\end{align}
The rest of the scalars and one-forms can be expressed in a similar fashion.

We note that the general solution in this orbit cannot be written in the Gibbons-Hawking form.  For example, this orbit contains the extremal non-rotating Rasheed-Larsen solution \cite{Rasheed:1995zv, Larsen:1999pp},
whose five-dimensional lift cannot be written as a time fibration over a hyper-K\"ahler
base. Though, the five-dimensional lift  of some members of this orbit can be written
as a time fibration over a hyper-K\"ahler
base. These geometries have been dubbed `almost BPS black holes' \cite{Goldstein:2008fq}. An important open problem is to find the `almost BPS' subsector of the  $\mathcal{O}_{4K}'$ orbit.

Since the analysis in
 this section is also valid for $n > 0$, the discussion above
also hold for the $\mathcal{O}_{3K}'$ orbit. However, it is
easy to see that this orbit is unphysical---a generic solution in this
orbit contains singularities outside or on the horizon. Since
 $\Diamond \ge 0$ for the $\mathcal{O}_{3K}'$ orbit, the warp factor
  (\ref{warpO4}) approaches negative infinity as
$r$ goes to zero when the quartic invariant is strictly positive. (It becomes zero at $r=0$ when $\Diamond = 0$.) This implies that the warp factor  becomes
zero at least at one point on or outside
the horizon. At this point the
spacetime is singular.
{The singularity is not a coordinate singularity but a genuine curvature singularity. As discussed in section \ref{sec:O3}, this can be most easily seen by looking at  curvature invariants. For example, the four-dimensional curvature invariant $R_{\mu \nu \rho \sigma} R^{\mu \nu \rho \sigma}$ diverges as $\phi_2$ tends to minus infinity. Thus, we conclude that the $\mathcal{O}_{3K}'$ orbit is unphysical.}

\subsubsection{Two parameter
$\mathrm{KKM}-\overline{\mathrm{M2}}^3$}\label{ssec:barM2KKM}
The first example we discuss is a static non-BPS electrically charged
black hole sitting at the center of the Taub-NUT. The solution takes
the `almost BPS' form
\begin{align}
ds^2 &= -f^2 dt^2 + f^{-1}\left[H^{-1} (dz +P \cos \theta d\phi)^2 +
H ds^2_3\right],\\
A &= -\sqrt{3} \left( 1-f\right) dt,
\end{align}
where
\begin{align}
f^{-1} &= 1+ \frac{Q}{r}, &  H &= 1 + \frac{P}{r},&P&> 0, &Q &> 0.
\end{align}
We use $P$ to denote the KKM charge. This solution is simply obtained
by flipping the sign of the Maxwell potential
 as compared to the
supersymmetric $\rom{M2^3-KKM}$ solution discussed in section
\ref{ssec:M2KKM}. The non-zero scalars for this solution are
\begin{align}
\phi_1 &=\frac{\sqrt{3}}{2}  \log \left( fH\right), & \phi_2
&=\frac{1}{2}\log\left(H f^{-3}\right),\nn\\
\chi_3 &= - \sqrt{3}\left(1-f\right), & \chi_5 &= \left(1-H^{-1}\right).
\end{align}
The charges are
\begin{align} \label{Mcharg2}
\Sigma &= \frac{P- Q }{2}, & M &= \frac{P+3  Q }{2}, \nn \\
Q_2 &= - Q, & P_1&= P.
\end{align}
Note the minus sign in $Q_2$, which yields that  the quartic invariant for the solution
is negative, $\Diamond = - 4P Q^3$, and therefore the solution is not supersymmetric.
One obtains this solution by setting
\begin{align}
 m &= Q, & d &= \frac{1}{3}\left( 3 + \frac{P}{Q}\right),
& \alpha &= \beta = 0,
\end{align}
in the parameterization
\eqref{paraO4M}--\eqref{paraO4Xi}
of the $\mathcal{O}_{4K}'$ orbit.

\subsubsection{Extremal non-rotating Rasheed-Larsen solution}
\label{eqn:nonsusy}
The  $\mathcal{O}_{4K}'$ orbit also contains the pure gravity extremal
non-rotating Rasheed-Larsen solution \cite{Rasheed:1995zv,
Larsen:1999pp}. The solution takes the form
\be
ds^2 = \frac{H_2}{H_1} (dz+ \cA)^2 - \frac{r^2}{H_2} dt^2 +
\frac{H_1}{r^2} \left(dr^2 + r^2 d\theta^2 + r^2 \sin^2\theta
d\phi^2\right),
\ee
where
\begin{align}
H_1 &= r^2 + r p + \frac{p}{p+q} \, \frac{pq}{2}, &  H_2 &= r^2 + r
q + \frac{q}{p+q} \, \frac{pq}{2},\nn\\
\cA_t &= \frac{q}{H_2} \sqrt{\frac{q}{p+q}}\, \left(r +
\frac{p}{2}\right), &  \cA_\phi &=p \sqrt{\frac{p}{p+q}} \, \cos
\theta,
\end{align}
with $p, q > 0$.
The non-zero scalars for this solution are
\begin{align}
\phi_1 &=\frac{1}{2} \sqrt{3} \log \frac{H_1}{H_2}, &
\phi_2 &=\frac{1}{2}\log \frac{H_1 H_2}{r^4}, \nn\\
\chi_1 &= \frac{1}{2} q \sqrt{\frac{q}{p+q}} \frac{p+2 r}{H_2}, &
\chi_5 &= \frac{1}{2} p \sqrt{\frac{p}{p+q}} \frac{q+2 r}{H_1},
\end{align}
and
\be
\chi_6 = \frac{(p q)^{3/2}}{2 (p+q) H_1}.
\ee
The non-zero charges are
\begin{align} \label{charges-nonsusy}
\Sigma &= \frac{p-q}{2}, & M &= \frac{p+q}{2},& Q_1 &= q
\sqrt{\frac{q}{p+q}}, & P_1&= p \sqrt{\frac{p}{p+q}}.
\end{align}
The quartic invariant for this solution evaluates to
\begin{align}
\Diamond = - \frac{p^3q^3}{(p+q)^3},
\end{align}
exhibiting that this solution is not supersymmetric. The solution can be obtained by setting
\begin{align}
m &= \frac{\sqrt{p q}}{2},&
d&= \frac{2}{3}\left( \frac{p+q}{ \sqrt{p q}}\right), &
\alpha &= \arcsin\frac{p-q}{p+ q},&
\beta &= -\arcsin\sqrt{\frac{p}{p + q}},
\end{align}
in the parameterization
\eqref{paraO4M}--\eqref{paraO4Xi}
of the $\mathcal{O}_{4K}'$ orbit.

\subsubsection{Three parameter $\rom{M2^3-M5^3-P}$ non-supersymmetric family}
\label{sec:nonsusystring}
A three parameter family of \emph{non-supersymmetric} black strings in minimal supergravity with all three independent $\rom{M2^3-M5^3-P}$ charges is also contained in $\mathcal{O}_{4K}'$ orbit. The solution is most conveniently parameterized by setting
\be
m = \frac{1}{2} \sqrt{q^2+Q^2},\quad d = \frac{2 \Delta }{\sqrt{q^2+Q^2}}, \quad \alpha  = \frac{\pi}{2} - \beta, \quad  \beta = \arccos\frac{Q}{\sqrt{q^2+Q^2}},
\ee
in the general expressions for the $\mathcal{O}_{4K}'$ orbit. The metric and Maxwell potential
after the coordinate transformation $z \to z +t$ in the general form of the $\mathcal{O}_{4K}'$ orbit are given by
\begin{align}
ds^2 &= V^2 ds^2_3 + 2 V^{-1} dt dz + V^{-4} g dz^2, \label{stringm2m5p}  \\
A &= - \frac{\sqrt{3}}{2} q \cos \theta d \phi - \frac{ \sqrt{3} Q}{r}\left(1+ \frac{q}{4r} \right) V^{-2}   dz,
\end{align}
where
\begin{align}
V &= 1 + \frac{q}{2r}, \\
g &= 1+\frac{3 \Delta }{r}-\frac{3\left(q^2-3 \Delta  q+Q^2\right) }{2 r^2} + \frac{9 q^2 \Delta -4 q \left(q^2+Q^2\right) }{4r^3} -\frac{\Diamond}{4r^4}.
\end{align}
The quartic invariant for this solution is
\be
\Diamond = - \frac{3}{4} q^2 \left(2 \Delta  q - q^2 - Q^2\right).
\ee
We restrict ourselves to the parameter range $Q \ge 0, q > 0$ and $  \Delta > \frac{q^2 + Q^2}{2q}$. In this range the quartic invariant is negative definite. The non-zero charges are
\begin{align}
& Q_2 = Q,& P_2 & = \frac{q}{2},& Q_1&= \frac{3}{2} (q-2 \Delta ),\nn\\
&M = \frac{3 \Delta }{2}, & \Sigma &= q-\frac{3 \Delta }{2}, &\Xi &= -Q.
\end{align}
Calculating the pressure density of the string \eqref{stringm2m5p}---the $T_{zz}$ component of the ADM stress tensor (see e.g. section 2.1 of \cite{Myers:1999psa})---we find
\be
T_{zz} = M + 3 \Sigma = 3 (q - \Delta).
\ee
When $\Delta = q$ the string becomes pressureless. The blackfold approach \cite{Emparan:2009cs}  strongly suggests that the  $\Delta = q$ string describes the infinite radius limit of a smooth black ring with these charges.
This black ring might be contained in the conjectured five parameter non-extremal black ring \cite{Elvang:2004xi} in this theory. This point deserves further investigation.


\section{Nilpotency and the Gibbons-Hawking form}
\label{sec:extremalsolutions}
We now briefly discuss how our analysis fits in with the approach of \cite{Gauntlett:2002nw}. It was shown in  \cite{Gauntlett:2002nw} that the most general supersymmetric solution of five-dimensional minimal supergravity with Gibbons-Hawking base space can be written in terms of four harmonic functions, $H, K, L$ and $M$ on $\mathbb{R}^3$. The functions $f$ and $\omega_5$ appearing in the solution take the form
\begin{equation}
\label{eqn:fharmonic}
f^{-1} = K^2 H^{-1}+L
\end{equation}
and
\begin{equation}
\label{eqn:wharmonic}
\omega_5 = H^{-2}K^3+\frac{3}{2} H^{-1}K L + M.
\end{equation}
Solutions with non-vanishing $\omega_5$ generically describe rotating, or boosted, five-dimensional spacetimes.

Using
(\ref{eqn:fsquareshifted})--(\ref{eqn:B2}) one can easily calculate the sigma model fields in terms of theses harmonic functions, and from the asymptotic expansion of these fields the general charge matrix can be obtained.
In order to make connection with our approach we  restrict the harmonic functions so that the five-dimensional solutions correspond to static spherically symmetric
asymptotically flat black holes in four dimensions after dimensional reduction over $z$. In particular, this requires that not only $\phi_1$, $\phi_2, \chi_1, \ldots, \chi_6$
vanish as $r \rightarrow \infty$ but also $\omega_3$
vanish identically. This imposes certain non-trivial restrictions on the general form of the harmonic functions. After taking these restrictions into account
one sees that the general charge matrix is indeed nilpotent.

In addition, one finds that
the resulting general charge matrix $\cQ$
commutes with a nilpotent element that belongs to
the $\mathrm{G}_{2(2)}$-orbit $\mathcal{O}_1$.
This implies that $\cQ$ is either in $\mathcal{O}_1$, $\mathcal{O}_2$,
$\mathcal{O}_{3K}$ or $\mathcal{O}_{4K}$.
As noted in section \ref{sec:O3},
solutions in the $\mathcal{O}_{4K}$ orbit are unphysical since they are
not regular outside the horizon in four dimensions.
Discarding these unphysical solutions, we conclude
that all static extremal asymptotically flat black holes  that can be obtained via dimensional reduction of solutions in the Gibbons-Hawking form of five-dimensional minimal supergravity are supersymmetric and belong to one of the $\tilde K$-orbits $\mathcal{O}_1$, $\mathcal{O}_2$ and $\mathcal{O}_{3K}$.


\subsection*{Acknowledgements}
We would like to thank Iosif Bena, Guillaume Bossard,  Borun D.~Chowdhury,
 Gianguido Dall'Agata,
 Roberto Emparan, Axel Kleinschmidt, Hermann Nicolai, and Clement Ruef for discussions.
 Our work is partially supported by IISN - Belgium (conventions 4.4511.06
and 4.4514.08) and by the Belgian Federal Science Policy Office through the
Interuniversity Attraction Pole P6/11.


\appendix

\setcounter{equation}{0}

\section{Non-linear sigma model for $\mathrm{G}_{2(2)} / \tilde{K} $}
\label{sec:SigmaModel}
Here we give a brief outline  of the sigma model
construction following \cite{Compere:2009zh}. A more detailed
discussion can be found in \cite{Compere:2009zh, Figueras:2009mc}.
Denoting the field strengths associated with  $B_{1}$ and $B_{2}$ by $\widetilde H_1$ and $\widetilde H_2$, respectively,
we define
\begin{align}
\widetilde H_1&=d B_1+ \Atwo\wedge d\Azero,
\end{align}
and
\begin{align}
\widetilde H_2&= d B_2- d\axone\wedge (B_1- \Azero \Atwo)- d\axtwo\wedge \Atwo.
\end{align}
In order to see the full hidden symmetry, the next step is to define
the three axions  $\chi_4$, $\chi_5$, and $\chi_6$ dual to the one-forms $\gauge$, $\Aone$, and $\Atwo$, respectively.
This is most conveniently done
by introducing one-form field strengths
$G_{4}$, $G_{5}$ and $G_{6}$ for the three axions:
\begin{align}
G_{4}&  \equiv e^{-\vec\alpha_4 \cdot \vec\phi}\star_3 \widetilde
H_2=  d\chi_4+\frac{1}{\sqrt 3} (\chi_2  d\chi_3 - \chi_3
d\chi_2), \nonumber \\
G_{5} &\equiv e^{-\vec\alpha_5 \cdot \vec\phi} \star_3  \widetilde
H_1 =  d\chi_5 - \chi_2  d\chi_4 + \frac{1}{3\sqrt{3}}
\chi_2 (\chi_3  d\chi_2 - \chi_2  d\chi_3),
\nn\\
G_{6} & \equiv-e^{-\vec\alpha_6 \cdot \vec\phi}\star_3 d\omega_3 =
d\chi_6- \chi_1  d\chi_5+ (\chi_1\chi_2 - \chi_3) d\chi_4\nonumber\\
&\qquad\qquad\quad\quad \qquad \quad \: + \:  \frac{1}{3\sqrt{3}} (-\chi_1\chi_2+\chi_3)
(\chi_3  d\chi_2 - \chi_2  d\chi_3).\label{eqn:dualfields}
\end{align}
The three-dimensional Lagrangian then can be written in terms of these scalars, $\phi_1$, $\phi_2$, $\chi_1,\dots,\chi_6$ as
\begin{align}
\cL_3 &= R \star_3 1- \half \star_3 d\vec\phi\wedge d\vec\phi + \half
e^{\vec\alpha_1 \cdot \vec\phi} \star_3  d\chi_1 \wedge d \chi_1 -
\half  e^{\vec\alpha_2 \cdot \vec\phi}\star_3  d\chi_2\wedge
d\chi_2\nonumber\\
&\quad\, + \: \half e^{\vec\alpha_3 \cdot \vec\phi}\star_3  (d\chi_3-\chi_1
d\chi_2)\wedge (d\chi_3-\chi_1 d\chi_2) + \half e^{\vec\alpha_4 \cdot
\vec\phi}\star_3 G_{4} \wedge  G_{4} \nonumber\\
&\quad\, + \:  \half  e^{\vec\alpha_5 \cdot \vec\phi}\star_3  G_{5} \wedge
G_{5} - \half e^{\vec\alpha_6 \cdot \vec\phi}\star_3  G_{6} \wedge
G_{6},\label{Lcoset_implicit}
\end{align}
where the six doublets  $\vec\alpha_1, \dots, \vec\alpha_6$ are the six
positive roots associated with the root vectors $e_1, \ldots, e_6$ of
$\mathfrak{g}_{2(2)}$.

The Lagrangian
(\ref{Lcoset_implicit}) can also be written as
\beq \cL_3 = R\star_3 1 + \cL_{\rom{scalar}} \, , \eeq
where $\cL_{\rom{scalar}}$ is the Lagrangian of a non-linear
sigma model for the coset $\mathrm{G}_{2(2)}/\tilde{K}$, with $\tilde{K} =
\mathrm{SO}_0(2,2)$. We can write a coset representative
$\cV$ for the coset $\mathrm{G}_{2(2)}/\tilde{K}$ in the Borel gauge by
exponentiating the Cartan elements and positive root vectors of
$\mf{g}_{2(2)}$ with the dilatons and axions as their coefficients. The representative
\begin{equation}
\mathbb \cV = e^{\frac{1}{2 \sqrt{3}}\phi_1 h_2 + \half \phi_2 \left(
h_2 + 2h_1 \right) } e^{\chi_1 e_1 }e^{-\chi_2 \frac{e_2}{\sqrt{3}}
+\chi_3 \frac{e_3}{\sqrt{3}}} e^{\chi_6 \frac{e_6}{6}} e^{\chi_4
\frac{e_4}{\sqrt{12}} -\chi_5 \frac{e_5}{6}},
\label{coset}
\end{equation}
reproduces
the Lagrangian (\ref{Lcoset_implicit}) \cite{Compere:2009zh}.
The matrix $\cM$ \eqref{M} is then given by
\beq
\cM = (\cV^{\sharp})  \cV, \label{MM}
\eeq
where $\sharp$ stands for the generalized transposition
\begin{align}
x^\sharp & \equiv -  \tau (x),
\label{generalizedtransposition}
\end{align}
for all $x \in \mf{g}_{2(2)}$, with the involution $\tau$ given in \eqref{invol1} and \eqref{invol2}. One uses (\ref{MM}) to derive the space-time fields from a geodesic.


\bibliographystyle{utphys}

\providecommand{\href}[2]{#2}\begingroup\raggedright\endgroup

\end{document}